\documentclass[journal,twoside,web]{ieeecolor}
\usepackage{generic}

\usepackage{cite}
\usepackage{amsmath,amssymb,amsfonts}
\usepackage{algorithmic}
\usepackage{graphicx}
\usepackage{algorithm,algorithmic}

\usepackage{caption}
\usepackage{subcaption}
\usepackage{xcolor} 
\usepackage{epsfig} 
\usepackage{bbm, bm} 
\usepackage{mathrsfs}
\usepackage{anyfontsize}

\usepackage{tikz}
\usetikzlibrary{intersections}
\usetikzlibrary{backgrounds}
\usetikzlibrary{arrows.meta}
\usetikzlibrary{backgrounds}
\tikzset{every node/.style={scale=3.2}}

\usepackage{pgfplots}
\usepgfplotslibrary{patchplots}
\usepgfplotslibrary{fillbetween}
\pgfplotsset{%
     layers/standard/.define layer set={%
         background,axis background,axis grid,axis ticks,axis lines,axis tick labels,pre main,main,axis descriptions,axis foreground%
     }{
         grid style={/pgfplots/on layer=axis grid},%
         tick style={/pgfplots/on layer=axis ticks},%
         axis line style={/pgfplots/on layer=axis lines},%
         label style={/pgfplots/on layer=axis descriptions},%
         legend style={/pgfplots/on layer=axis descriptions},%
         title style={/pgfplots/on layer=axis descriptions},%
         colorbar style={/pgfplots/on layer=axis descriptions},%
         ticklabel style={/pgfplots/on layer=axis tick labels},%
         axis background@ style={/pgfplots/on layer=axis background},%
         3d box foreground style={/pgfplots/on layer=axis foreground},%
     },
 }
\pgfplotsset{compat = 1.16}
\pgfplotsset{every node/.style={scale=3.3}}

\usepackage{ulem}
\usepackage{hyperref}
\hypersetup{hidelinks=true}
\usepackage{textcomp}

\usepackage{balance}

\newcommand{\useTIKZ}{1} 

\newcommand {\st}{\text{\footnotesize $(t)$}}
\newcommand {\so}{\text{\footnotesize $(0)$}}
\newcommand {\sta}[1]{\text{\footnotesize $(#1)$}}

\newcommand {\vbeta}{\vec{\beta}}

\newcommand {\cB}{{\mathcal{B}}}

\newcommand {\R} {{\rm I\kern-2.5pt R}}

\newtheorem{lemma}{Lemma}

\newtheorem{proposition}{Proposition}

\newtheorem{corollary}{Corollary}
\newtheorem{theorem}{Theorem}
\newtheorem{remark}{Remark}
\newtheorem{example}{Example}
\newtheorem{definition}{Definition}

\newcommand{\beqa}{\begin{eqnarray}}
\newcommand{\eeqa}{\end{eqnarray}}
\newcommand{\beqan}{\begin{eqnarray*}}
\newcommand{\eeqan}{\end{eqnarray*}}
\newcommand{\beq}{\begin{equation}}
\newcommand{\eeq}{\end{equation}}

\newcommand{\bfl}{\begin{flushleft}}
\newcommand{\efl}{\end{flushleft}}
\newcommand{\bgnv}{\begin{verbatim}}
\newcommand{\ednv}{\end{verbatim}}

\newcommand{\myarr}{\begin{array}{lll}}

\newcommand{\bitem}{\begin{itemize}}
\newcommand{\eitem}{\end{itemize}}
\newcommand{\benum}{\begin{enumerate}}
\newcommand{\eenum}{\end{enumerate}}

\def\QED{~\rule[-1pt]{5pt}{5pt}\par\medskip}

\newcommand{\change}[1]{{{\color{black}{#1}}}}

\newcommand{\orcid}[1]{\href{https://orcid.org/#1}{\textcolor[HTML]{A6CE39}{\aiOrcid}}}

\DeclareMathOperator*{\argmax}{arg\,max}

\begin{document}
\title{Epidemic Population Games And \\ Perturbed Best Response Dynamics}
\author{Shinkyu Park, \IEEEmembership{Member, IEEE}, Jair Certorio, \IEEEmembership{Student Member, IEEE}, \\ Nuno C. Martins, \IEEEmembership{Senior Member, IEEE}, Richard J. La
    \thanks{
        Corresponding author: N.~C.~Martins. The work of Park was supported by
        funding from King Abdullah University of Science and Technology (KAUST). This work was
        supported by AFOSR Grant FA9550-19-1-0315 and NSF Grant 2135561.  }
    \thanks{Shinkyu Park is with the CEMSE, at King Abdullah University of Science and Technology. (e-mail: shinkyu.park@kaust.edu.sa). }
    \thanks{Jair Certorio, Nuno C. Martins, and Richard J. La are with the Dept. of ECE and ISR, at the University of Maryland, at College Park. (e-mails: certorio@umd.edu, nmartins@umd.edu, hyongla@umd.edu).}
}

\maketitle

\begin{abstract}                       
This paper proposes an approach to mitigate epidemic spread in a population of strategic agents by encouraging safer behaviors through carefully designed rewards. These rewards, which adapt to the evolving state of the epidemic, are ascribed by a dynamic payoff mechanism we seek to design. We use a modified SIRS model to track how the epidemic progresses in response to the agents' strategic choices. By employing perturbed best response evolutionary dynamics to model the population's strategic behavior, we extend previous related work so as to allow for noise in the agents' perceptions of the rewards and intrinsic costs of the available strategies. Central to our approach is the use of system-theoretic methods and passivity concepts to obtain a Lyapunov function, ensuring the global asymptotic stability of an endemic equilibrium with minimized infection prevalence under budget constraints. We leverage the Lyapunov function to analyze how the epidemic’s spread rate is influenced by the time scale of the payoff mechanism’s dynamics. Additionally, we derive anytime upper bounds on both the infectious fraction of the population and the instantaneous cost a social planner must incur to control the spread, allowing us to quantify the trade-off between peak infection prevalence and the corresponding cost. For a class of one-parameter perturbed best response models, we propose a method to learn the model's parameter from data.
\end{abstract}


\section{Introduction}

Recent studies~\cite{MARTINS2023111016,certorio_epidemic_2022} have used system-theoretic methods to obtain dynamic rewards promoting safer behaviors during an epidemic, aiming to lessen infection prevalence in a large population. The results in~\cite{MARTINS2023111016,certorio_epidemic_2022} apply to a wide range of strategic decision-making behaviors but assume that the agents have precise knowledge of the rewards and costs of the available strategies. Unlike these studies, here we accept that the agents' choices are based on rough estimates of the strategies' net rewards (rewards minus intrinsic costs for adopting the strategies), leading us to explore the so-called \textit{Perturbed Best Response (PBR) rules} to capture the unavoidable noise in the agents' perceptions.

Our work stands on two pillars: First, we use a modified susceptible-infectious-recovered-susceptible (SIRS) epidemic model \cite{pastor-satorras_epidemic_2015} to track disease spread. The agents' strategic choices affect the infection rate denoted as $\cB\st$. The agents choose from a range of strategies $\{1,\ldots,n\}$, $n \geq 2$, each affecting $\cB\st$. The strategies' net rewards, calculated by rewards minus intrinsic costs, guide the decisions the agents make. This decision-making is described by an evolutionary dynamics model, which assumes that the agents can review and change their strategies over time according to a PBR rule~\cite{Hofbauer2002On-the-global-c, Hofbauer2007Evolution-in-ga} (see~\S\ref{sec:pbr_learning_rule}). Second, we seek to design a payoff mechanism for assigning rewards that nudge the population's choices towards safer strategies reducing infection rates, with the constraint of keeping rewards affordable in the long run. The payoff mechanism's state is coupled to the epidemic model's state, creating what we call an \textit{Epidemic Population Game (EPG)} (see \S\ref{sec:pbr_learning_rule}).

\change{
The design methods proposed in \cite{MARTINS2023111016, certorio_epidemic_2022} are not suitable for our context, as they assume the absence of noise. While our approach also builds on system-theoretic passivity as in \cite{MARTINS2023111016, certorio_epidemic_2022}, it requires substantial modifications to the stability proofs and the characterization of the set of endemic equilibria with minimized infection prevalence, which here depends on the noise distribution. For an important subset of PBR rules, which we will describe later, this distribution is governed by a single parameter that can be learned from data. Moreover, we establish local exponential stability for a representative class of PBR dynamics.  Additionally, we explore the impact of the time scale governing the dynamics of the payoff mechanism on the upper bound of the infected population. By incorporating this time scale parameter into the design, it becomes an additional lever that a social planner can adjust to mitigate the spread of the endemic.
}

\subsection{Evolutionary Dynamics Model}
\label{subsec:EDM}

Each agent follows one strategy at a time, which it can revise repeatedly. A payoff vector $p\st$ in $\mathbb{R}^n$ whose entries quantify the net rewards of the available strategies influences the revision process. Typically, the agents are more likely to choose strategies with higher payoffs. \change{Namely, we define
\begin{align}
    \label{eq:payoffEq}
    p\st : =  r\st - c (I\st),
\end{align} 
where $I\st$ represents the infectious portion of the population, $c(I\st) = (c_1(I\st), \cdots, c_n(I\st))$ is a vector whose $\ell$-th entry $c_\ell (I\st)$ is the inherent cost of the $\ell$-th strategy, which depends continuously on $I\st$ and decrease as $I\st$ increases, and $r\st = (r_1\st, \cdots, r_n\st)$ is a reward vector meant to incentivize the adoption of safer (costlier) strategies, where $r_\ell\st$ is the $\ell$-th strategy's reward.}

Rather than focusing on what each strategy may represent (see \cite[Remark~1]{MARTINS2023111016}), in our framework, we assume that a vector  $\vbeta$ in $\mathbb{R}_{>0}^n$ is given whose $\ell$-th entry $\vbeta_\ell$ quantifies the effect of the $\ell$-th strategy towards $\cB\st$ according to
\begin{equation}
    \label{eq:betaaverage}
    \cB\st = \vbeta' x\st, \quad t \geq 0,
\end{equation} where $x\st$ is the so-called {\it population state} taking values in the standard simplex $\mathbb{X}$ defined below and whose $\ell$-th entry $x_\ell\st$ represents the proportion of the population adopting the $\ell$-th strategy at time $t$.
\begin{equation*}
    \mathbb{X}:= \left \{x \in [0,1]^n \ \big | \ \textstyle\sum_{i=1}^n x_i =1 \right \}.
\end{equation*}

Following the standard approach in~\cite[\S4.1.2]{Sandholm2010Population-Game}, the following {\it Evolutionary Dynamics Model (EDM)} governs the dynamics of $x\st$ in the large-population limit:
\begin{subequations}
    \begin{equation}\tag{EDMa} \label{eq:EDM-DEF} \dot {x}\st = \mathcal{V} ( x\st,p\st ), \quad t\geq 0,
    \end{equation}
    where the $i$-th component of $\mathcal{V}$ is specified as
    \begin{align}
        \nonumber \mathcal{V}_i ( x\st,p\st ) := & \underbrace{- \textstyle \sum_{j=1}^{n} x_i\st \mathcal{T}_{i j} ( x\st,p\st ) }_{\text{\footnotesize outflow switching from strategy $i$}} \\ \tag{EDMb} \label{eq:EDMfromProtocol}  & \underbrace{+ \textstyle \sum_{j=1}^{n}  x_j\st \mathcal{T}_{ji} ( x\st,p\st )}_{\text{\footnotesize inflow switching to strategy $i$}}.
    \end{align}
\end{subequations}
\change{A Lipschitz continuous map $\mathcal{T}: \mathbb{X} \times \mathbb{R}^{n} \rightarrow [0, 1]^{n \times n}$ is referred to as the learning rule (or revision protocol). Specifically, given the current state $x\st$ and the payoff vector $p\st$, $\mathcal{T}_{ji}(x\st, p\st)$ represents the probability that an agent will switch from its current strategy~$j$ to strategy~$i$ when given the opportunity. As discussed in \cite[Part II]{Sandholm2010Population-Game} and \cite[\S 13.3-13.5]{Sandholm2015Handbook-of-gam}, which provide an in-depth exploration of various learning rules, the opportunity for strategy revision for each agent is modeled by an i.i.d. Poisson process.}


\subsection{PBR Learning Rule and Epidemic Population Game} \label{sec:pbr_learning_rule}

In contrast to earlier work in~\cite{MARTINS2023111016,certorio_epidemic_2022,certorio_CDC23}, here we adopt the class of PBR learning rules~\cite{Hofbauer2002On-the-global-c, Hofbauer2007Evolution-in-ga} as a way to incorporate noise in the agents' perceived payoffs. Specifically, we consider
\begin{multline} \label{eq:perturbed_revision_protocol}
    \mathcal T_{ji} (x,p) = C_i(p) := \mathbb P \Big( p_i + v_i \geq \max_{1 \leq \ell \leq n} (p_\ell + v_\ell ) \Big), \\ \forall j \in \{1, \cdots, n\},
\end{multline}
where $v_1, \cdots, v_n$ are random variables admitting a positive joint probability density function over $\mathbb R^n$. 
In the following example, we discuss a learning rule that has been widely studied in the literature.

\begin{example} 
The logit learning rule \cite{Hofbauer2007Evolution-in-ga} is specified as
    \begin{align} \label{eq:logit_learning_rule}
        C_i(p) \underset{\text{\tiny logit}}{=} \frac{e^{p_i / \mu}}{\sum_{\ell =1} ^n e^{p_\ell / \mu}},
    \end{align}
    where $\mu > 0$ quantifies the noise intensity. Namely, $v_1, \cdots, v_n$ are i.i.d. random variables characterized by $\mathbb{P}(v_i \leq \zeta) = e^{-e^{-\zeta / \mu - \bar{\zeta}}}$, where $\bar{\zeta}$ is Euler's constant.
    In the noise-free limit as $\mu$ tends to zero, $C$ will tend to the best response learning rule.
\end{example}

According to \cite{Hofbauer2002On-the-global-c}, we can represent the choice function as\footnote{Since $Q$ satisfies \eqref{eq:perturbation_conditions}, according to the analysis presented in \cite{Hofbauer2002On-the-global-c}, the maximization in \eqref{eq:choice_function_p} admits a unique solution.}
\begin{align} \label{eq:choice_function_p}
    C(p) = (C_1(p), \cdots, C_n(p)) = \argmax_{z \in \mathrm{int} (\mathbb X)} ( z' p \!-\! Q (z) ),
\end{align}
where $Q: \mathrm{int}(\mathbb X) \to \mathbb R$ is called the \textit{admissible payoff perturbation}, which is twice continuously differentiable and satisfies the following conditions:
\begin{subequations} \label{eq:perturbation_conditions}
    \begin{align}
        \tilde z' \nabla^2 Q(z) \tilde z                      & > 0, ~ z \in \mathbb X, \tilde z \in T\mathbb X \setminus \{0\} \label{eq:perturbation_conditions_a}               \\
        \lim_{z_{\text{\tiny min}} \to 0} \| \nabla Q(z) \|_2 & = \infty, ~ \text{ where } z_{\text{\tiny min}} = \min_{1 \leq i \leq n} z_i, \label{eq:perturbation_conditions_b}
    \end{align}
\end{subequations}
where $T \mathbb X$ is the tangent space of $\mathbb X$. The logit learning rule is obtained via the perturbation function $Q$ defined as $Q(x) = \mu \sum_{i=1}^n x_i \ln x_i$, with $\mu >0$.

When (EDM) is defined by the choice function \eqref{eq:choice_function_p}, we refer to it as the PBR EDM. Note that when the PBR EDM reaches its equilibrium state, i.e., $\mathcal V(x^\ast, p^\ast) = 0$, the population state $x^\ast$ satisfies $x^\ast = C(p^\ast)$.


We adopt the EPG formalism from \cite{MARTINS2023111016} as follows.
\begin{align} \tag{EPGa}
    \dot{I}\st & = \big ( \cB \st (1-I\st-R\st) -\sigma) I\st \\ \tag{EPGb}
    \dot{R}\st & = \gamma I\st - \omega R\st                  \\ \tag{EPGc}
    \dot{q}\st & = G (I\st,R\st,x\st,q\st)                    \\ \tag{EPGd}
    r\st       & = H(I\st,R\st,x\st,q\st),
\end{align} 
where $I\st$, $R\st$, and $S\st:=1-I\st-R\st$ take values in $[0,1]$ and represent the proportions of the population which are infectious, have recovered, and are susceptible to infection at time $t$, respectively. Here, (EPGa,b) is a normalized SIRS model with ${\sigma:= \gamma + \theta}$ and ${\omega:=\psi + \theta}$, where  $\gamma$ and $\psi$ denote the daily recovery rate and the daily rate at which recovered individuals become susceptible (due to waning immunity), respectively, and $\theta$ is the daily birth rate. Additionally, (EPGc,d) is a payoff mechanism we seek to design, where $r\st$ appears in~(\ref{eq:payoffEq}) and $q\st$ belongs to $\mathbb{R}^m$ with $m \geq 1$.


\subsection{Problem Formulation and Paper Structure}
\label{sec:Problem}

\change{We order the entries of $\vbeta$ and $c (I)$~as follows:
\begin{equation*}
    \vbeta_i < \vbeta_{i+1} \text{ and } c_i (I) > c_{i+1} (I), ~ i \in \{1, \cdots, n-1 \}, ~ I \in [0,1].
\end{equation*} 
}
\noindent Henceforth, $\vbeta$ and $c:[0,1] \to \mathbb R^n$ satisfying the conditions above are assumed given.
We will use $\tilde{c} (I)$ defined below to specify cost constraints because for a planner seeking to promote the $i$-th strategy it suffices to offer incentives to offset the differential~$\tilde{c}_i (I)$.
\begin{equation} \label{eq:c_tilde}
    \tilde{c}_i (I):=c_i (I)-c_n (I), ~ i \in \{1, \cdots, n\}, ~I \in [0, 1].
\end{equation}

\change{
\begin{definition}  \label{def:betastar}  
Given a cost budget $c^* > 0$,
the optimal endemic transmission rate $\beta^*$ is determined as the minimum value of the following constrained optimization problem:
    \begin{align} \label{eq:betadef}
        &\min_{\beta \in (\sigma, \vec\beta_n), r \in \mathbb R_{\geq 0}^n, I \in [0, 1]} \beta \\
        &\begin{aligned}
            \text{subject to } & r' C(r-\tilde c (I)) \leq c^* \nonumber \\
            & \beta = \vbeta' C(r-\tilde c (I)) \\
            & I = \eta(1 - \sigma / \beta),
        \end{aligned}
    \end{align}
    where $C$ is the choice function defined in \eqref{eq:choice_function_p}.  
    
    Note that at the optimal solution $(\beta^\ast, r^\ast, I^\ast)$ for \eqref{eq:betadef}, $C(r^\ast-\tilde{c}(I^\ast))$ represents the population state at the equilibrium of the PBR EDM, where the infectious population $I^\ast$ is minimized while satisfying the budget constraint ${r^\ast}' C(r^\ast - \tilde{c}(I^\ast)) \leq c^\ast$. Additionally, we can infer that the optimal endemic transmission rate $\beta^\ast$ decreases as the cost budget $c^\ast$ increases, implying that with a higher $c^\ast$, the planner can further reduce $I^\ast$. Throughout the paper, we assume that an average transmission rate less than or equal to $\sigma$ is considered too costly for a social planner to sustain. Therefore, we focus on scenarios where the minimum $\beta^\ast$ is greater than $\sigma$.
    
\end{definition}

}

\vspace{.5em}

\noindent {\bf Main Problem:} We seek to obtain Lipschitz continuous $G$ and $H$ for which the following hold for any  $I\so$ in $(0, 1]$, $R\so$ in $[0,1-I\so]$, $x\so$ in $\mathbb{X}$, and $q\so$ in~$\mathbb{R}^m$:\footnote{\change{In our main results, specifically Corollary~\ref{corollary:stability}, the design of (EPGc) and (EPGd) guarantees that $r\st$ converges to $r^\ast$, as defined by the social planner in \eqref{eq:betadef}. It is important to note that $r^\ast$ is a vector with non-negative elements.}}
\begin{align} \tag{P1}
    \lim_{t \rightarrow \infty} (I,R,\cB)\st & = (I^*,R^*,\beta^*), \\ \tag{P2}  \limsup_{t \rightarrow \infty} r'\st x\st &\leq c^*  ,
\end{align} where, from Picard's Theorem, $\{(I,R,x,q)\st \ | \ t\geq 0\}$ is the unique solution of the initial value problem for the closed-loop model formed by~(EDM) and~(EPG). Here, the nontrivial endemic equilibrium for (EPGa,b)~is:
\begin{equation*}
    I^* : = \eta(1 - \tfrac{\sigma}{\beta^*}), ~ R^* : = (1-\eta) (1 - \tfrac{\sigma}{\beta^*}), ~ \eta:= \tfrac{\omega}{\omega+\gamma}.
\end{equation*}
\change{
}

We will seek $G$ and $H$ for which a Lyapunov function for the closed-loop model exists. We will do so not only to establish (P1) but, crucially, also to leverage the Lyapunov function to obtain anytime upper bounds for $I\st$. This is relevant because, as has been pointed out in studies~\cite{Godara2021A-control-theor,Sontag2021An-explicit-for} employing $\cB\st$ as a control variable, $I\st$ tends to significantly overshoot its endemic equilibrium $I^*$ when $I\sta{0}<I^*$, unless the control policy prevents it.

\noindent {\bf Paper Structure:} After the introduction, in \S\ref{sec:MotAndComp} we outline motivation for our work and provide a brief overview of related work. The paper's main section is \S\ref{sec:solution} where we describe $G$ and $H$ that constitute a solution to our Main Problem and present Theorem~\ref{theorem:stability} and Corollary~\ref{corollary:stability} stating the precise conditions under which (P1)-(P2) are guaranteed globally for the chosen $G$ and $H$ along with a method to select key parameters. In \S\ref{sec:solution}, we also discuss convergence rate, present an example, and explain how to construct anytime upper bounds for the prevalence of infections and how to learn from data the noise parameter specifying a class of choice functions when it is unknown a priori. In \S\ref{sec:sims}, we illustrate our methods via simulation for two scenarios.

\section{Motivation and Comparative Literature Review}
\label{sec:MotAndComp}

In this work, we investigate agent decision-making in epidemic processes where the agents' strategy selection is determined by PBR models. Our study is motivated by prevalent empirical observations, as exemplified in \cite{Mao2017}, which indicate that the Nash equilibrium is not an accurate predictor of human decision-making processes. We provide a review of prior studies on epidemic models and PBR models.

There are a number of recent studies that investigated the problem of managing an epidemic using control theory: di Lauro et al.~\cite{diLauro2021} and Sontag~\cite{Sontag2021An-explicit-for} studied the problem of identifying the optimal timing for non-pharmaceutical interventions (NPIs), e.g., quarantine and lockdowns, to minimize the peak infections. Al-Radhawi et al. \cite{Al-Radhawi2021} examined the problem of tuning NPIs to regulate infection rates as an adaptive control problem and investigated the stability of disease-free and endemic steady states. Godara et al. \cite{Godara2021A-control-theor} studied the problem of controlling the infection rate to minimize the total cost till herd immunity is attained as an optimal control problem subject to a constraint on the fraction of infectious population. But, these studies did not consider the strategic decision-making.

Several recent studies employed game theory, including evolutionary or population games, to study epidemic processes with strategic agents \cite{amaral_epidemiological_2020, Bauch2004, d'Onofrio2011, hota_game-theoretic_2019,kabir_evolutionary_2020, 10117559, Khazaei2021}. For instance,  \cite{amaral_epidemiological_2020} studied the effect of risk perception on whether individuals choose to self-quarantine or not, and how increased perceived risks could lead to multiple infection peaks. Khazaei et al.~\cite{Khazaei2021} adopted the SEIR epidemic model with the replicator dynamics to study the interplay between the underlying epidemic state and the behavioral response of a single population. They showed that as the disease prevalence changes over time, the level of public cooperation varies as well in response, which results in oscillations of infection level.

The dynamics of epidemic processes on networks have been studied extensively, e.g.,~\cite{pastor-satorras_epidemic_2015, nowzari_analysis_2016, Mei2017, Pare2020, paarporn_networked_2017}. Other studies also considered epidemics with multiple populations~\cite{alutto_multiple_2022, kuniya_global_2014}; each population represents a group of similar agents, a community or a geographic area (e.g., a city). The interactions among populations are often modeled using a graph, in which edge weights indicate the contact or interactions rates across different populations. We refer an interested reader to \cite{pastor-satorras_epidemic_2015, nowzari_analysis_2016} for a comprehensive literature survey. Considering multiple populations complicates the analysis of epidemic models, even without modeling agents' strategic interactions and leads to richer dynamics \cite{alutto_multiple_2022}. The impact of asymptomatic infections over complex networks has also been studied, along with seasonal transmission rate changes, e.g., a high tourist season or the start of a new school year \cite{stella_role_2022}.

Kuniya and Muroya \cite{kuniya_global_2014} studied a multi-group SIS model with population migration and established global convergence to the endemic equilibrium when the basic reproduction number exceeds one, and to the disease free equilibrium otherwise. These studies, however, assume fixed transmission rates that do not depend on the strategies chosen by the agents in different populations.
\\ \vspace{-0.12in}

The authors of \cite{MARTINS2023111016, certorio_epidemic_2022} introduced a new framework that combines the strategic decision-making process of agents (evolutionary dynamics) and a compartmental epidemic model (SIRS model) for a single population. This framework allowed them to design a dynamic payoff mechanism that ensures the convergence to an endemic equilibrium where the disease transmission rate is minimized subject to a budget constraint. A key contribution of these studies is that they provide {\em anytime} bounds on the peak infection, which are universal and hold for any protocol that meets certain assumptions. The framework was extended to two-population scenarios in \cite{certorio_CDC23}. However, these studies only considered learning rules that satisfy {\em Nash stationarity}. Although many learning rules are Nash stationary (e.g., impartial pairwise protocol), there are some well-known learning rules that are not Nash stationary (e.g., imitation and PBR dynamics).

Here, we extend the framework proposed in \cite{MARTINS2023111016, certorio_epidemic_2022} to study PBR dynamics in epidemic population games.
The PBR model (often referred to as the \textit{better response model}), which is rooted in bounded rationality and accounts for random noise in the payoff mechanism, has been extensively investigated in the economics literature \cite{CHEN199732, 10.2307/1061555, Harsanyi1973, Rosenthal1989, MCKELVEY19956, BLUME2003251, GOEREE2005349}. An earlier work relevant to the perturbed decision-making is \cite{Harsanyi1973} which presents a game formulation with randomly perturbed payoffs in finite-agent game settings. The mathematical analysis presented in the paper provides a rigorous approach to identifying the equilibrium, referred to as the \textit{Bayesian equilibrium} \cite{2f3e8fce-f2ca-35f7-af3c-454ab8ea0188}, of such a game with perturbed payoffs.

Rosenthal~\cite{Rosenthal1989} discussed the limitations of the Nash equilibrium in predicting outcomes of empirical studies in non-cooperative games. Motivated by such limitations, the author proposed a one-parameter better response model to capture the key characteristics in human decision-making: humans are equally likely to choose strategies that are equally effective. The study also provided concrete examples to illustrate the distinction between the equilibrium determined by the model and that of Nash's.

McKelvey and Palfrey~\cite{MCKELVEY19956} introduced a quantal choice model that includes the logit protocol as its special case and analyzed its equilibrium, referred to as the \textit{quantal response equilibrium (QRE)}. By examining data from various human decision-making experiments, they illustrated that QRE for the logit learning rule consistently provides a better prediction across these experiments compared to the Nash equilibrium. Additionally, the study reports variability in the selection of the optimal model parameter during the analysis of experimental data, indicating the need for developing a parameter estimation method. More in-depth treatment of the logit learning rule is presented in \cite{10.2307/1061555}. The authors explained why the Nash's approach to defining an equilibrium, which relies only on the signs of payoff differences, falls short in predicting the outcomes of the experiments adopted from \cite{15bd6604-fa38-3d3a-aa55-f40d669b2e78}.

Other related studies include \cite{BLUME2003251} which presents a model that captures the noise in agent decision-making to study the robustness of the \textit{stochastic stability}. \cite{GOEREE2005349} introduces the idea of \textit{stochastic potential} for a coordination game, which adds an effort cost to an ordinary potential function of the game, and draws the connection between the equilibrium maximizing the stochastic potential and the logit equilibrium. Using experimental data, the authors claim that such equilibrium is a better predictor of the experiment outcomes than that of Nash's.

\section{A Solution to Main Problem}
\label{sec:solution}
Recall the state equation of the PBR EDM given by
\begin{align} \label{eq:pbr_edm}
    \dot x \st & = \mathcal V(x\st, p\st) \nonumber                                                     \\
               & = \argmax_{z \in \mathrm{int} (\mathbb X)} ( z' p \st - Q (z) ) - x \st, \tag{PBR EDM}
\end{align}
where the payoff vector $p\st$ is defined by the reward $r\st$ and the intrinsic cost function $\tilde c: [0,1] \to \mathbb R_{\geq 0}^n$ as in \eqref{eq:payoffEq} and \eqref{eq:c_tilde}. We adopt the notion of $\delta$-passivity from \cite{Park2019From-Population} to analyze the stability of (PBR EDM) when it is interconnected with (EPG). (PBR EDM) is \textit{$\delta$-passive} in that there is a continuously differentiable function $\mathcal S: \mathbb X \times \mathbb R^n \to \mathbb R_{\geq 0}$ given by
\begin{align} \label{eq:delta_storage_function}
    \mathcal S(x,p) & = \max_{z \in \mathrm{int}(\mathbb X)} ( z' p - Q(z) ) - ( x' p - Q(x) ),
\end{align}
which satisfies
\begin{multline}
    \nabla_x' \mathcal S(x,p) \mathcal V (x,p) + \nabla_p' \mathcal S(x,p) u \leq u' \mathcal V (x,p), \\ \forall x \in \mathbb X, p \in \mathbb R^n, u \in \mathbb R^n.
\end{multline}
Note that $\nabla_x' \mathcal S(x,p) \mathcal V (x,p) \leq 0$ holds for all $x \in \mathbb X$, $p \in \mathbb R^n$, and $\mathcal S(x,p) = 0$ if and only if $\mathcal V(x,p) = 0$. We refer to $\mathcal S$ as the \textit{$\delta$-storage function} of (PBR EDM).

Following the similar idea as in \cite{MARTINS2023111016}, we define $G$ and $H$ as follows.
\change{
\begin{subequations} \label{eq:GH}
    \begin{align}
        G(I, R, x, q) & = \kappa \bigg( (\hat I - I) + \eta (\ln I - \ln \hat I) + \upsilon^2 (\bar\beta - \mathcal B) \nonumber \\
                      & \qquad \quad + \frac{\mathcal B}{\gamma} (R - \hat R)(1 - \eta - R) \bigg) \label{eq:G} \\
        H(I, R, x, q) & = q \vec \beta + \bar r + \tilde c(I) - \tilde c(\bar I), \label{eq:H}
    \end{align}
\end{subequations}
where $\bar I = \eta ( 1 - \sigma / \bar{\beta} )$, $\hat I = \eta ( 1 - \sigma / \mathcal B )$, $\hat R = (1 - \eta) ( 1 - \sigma / \mathcal B )$, $\mathcal B = \vec \beta' x$, and $\eta = \omega / (\omega + \gamma)$. The constant $\kappa$ represents the time scale of the dynamics for $q\st$, and its effect on the convergence of $I\st$ is discussed in \S\ref{sec:convergence_results}. The constants $\bar \beta \in (\sigma, \vec\beta_n)$ and $\bar r \in \mathbb R_{\geq 0}^n$ are the endemic transmission rate and stationary reward vector when $q\st$ vanishes as $t$ tends to infinity, respectively. These are design parameters of (EPG). The variables $\hat I$ and $\hat R$ constitute the equilibrium state of (EPGa) and (EPGb), and are considered as the \textit{reference} endemic variables. From this perspective, the parameter $\upsilon$ can be interpreted as the weight that determines the importance of the current transmission rate $\mathcal B$ attaining the desired value $\bar \beta$ compared to $(I, R)$ attaining $(\hat I, \hat R)$. 
}

\change{
The following are remarks on the choice of $G$ and $H$.
\begin{remark} \label{remark:choice_of_H}
    According to the dynamics for $q$ defined in \eqref{eq:GH}, $q$ is a signal adjusted by the social planner to control the portion $q\vec\beta$ in the reward vector \eqref{eq:H}. Specifically, when $q < 0$, the reward vector incentivizes more effective strategies -- those associated with lower values of $\vec\beta_i$. Additionally, $G$ in \eqref{eq:G} is defined such that $\frac{\partial G}{\partial \mathcal{B}} < 0$ holds. As a result, when agents switch to strategies that increase the average transmission rate (i.e., $\dot{\mathcal{B}} \st > 0$), the value of $G$ would decrease. If this trend persists, it will further lower the value of $q \st$, encouraging agents to adopt more effective strategies. Furthermore, using Lemma~\ref{lemma:c_increasing_function} (in Appendix~\ref{sec:proof_stability_theorem}), we can infer that $\mathcal B\st > \sigma, ~ \forall t \geq 0$, provided that $\mathcal B\so > \sigma$.

    If we want to ensure that the planner is offering rewards (non-negative incentives) to the population, the design of $H$ can be modified in the following way.
    \begin{multline}
      H(I, R, x, q) = q \vec \beta + \bar r + \tilde c(I) - \tilde c(\bar I) \\
      - \min_{1 \leq i \leq n} (q \vec \beta_i + \bar r_i + \tilde c_i(I) - \tilde c_i(\bar I) ) \mathbf 1. \label{eq:H_nonnegative_incentive}
    \end{multline}
    Since adding an equal value to each entry of the payoff vector $p$ does not affect the results of this work, we use \eqref{eq:H} for concise presentation, unless otherwise specified.
\end{remark}
}

\subsection{Optimal Design of $\bar r, \bar \beta$} \label{sec:convergence_results}
We begin by presenting the main convergence results.

\change{
\begin{theorem} \label{theorem:stability}
    Consider the closed-loop model consisting of (EPG) and (PBR EDM), with  the parameters $\bar \beta \in (\sigma, \vec\beta_n)$ and $\bar r \in \mathbb R_{\geq 0}^n$ given. Suppose $\mathcal B\st \geq \mathcal B_{\text{min}}, ~\forall t \geq 0$ holds for some $\bar\beta > \mathcal B_{\text{min}} > \sigma$; then the following statements hold:
    \begin{enumerate}
        \item $\lim_{t \to \infty} q\st = \bar q$ and $\lim_{t \to \infty} \mathcal B\st = \bar\beta$, where $\bar q$ is a unique solution to $\bar \beta = \vec\beta' C(\bar q \vec\beta + \bar r - \tilde c (\bar I))$ with $\bar I = \eta (1 - \sigma / \bar \beta)$, and

        \item $\lim_{t \to \infty} I\st = \eta (1 - \sigma / \bar \beta)$, $\lim_{t \to \infty} R\st = (1 - \eta) (1 - \sigma / \bar \beta)$, and $\lim_{t \to \infty} x\st = C(\bar q \vec\beta + \bar r - \tilde c (\bar I))$.
    \end{enumerate}
\end{theorem}
\vspace{.5em}
\begin{proposition} \label{prop:local_exponential_stability}
    When $C$ in (PBR EDM) is the logit learning rule \eqref{eq:logit_learning_rule} and $\upsilon$ in (EPGc) is sufficiently small, the closed-loop model, consisting of (EPG) and (PBR EDM), achieves local exponential stability.
\end{proposition}

The proofs are provided in Appendices~\ref{sec:proof_stability_theorem} and \ref{sec:proof_local_exponential_stability}.}
\vspace{.5em}
\change{
\begin{remark}
    The assumption $\mathcal B\st \geq \mathcal B_{\text{min}}, ~\forall t \geq 0$ is imposed to ensure that the term $\ln I\st - \ln \hat I\st$ in \eqref{eq:G} converges to zero as does $I\st - \hat I\st$. Note that if $\vec\beta_1 > \sigma$, this assumption is satisfied with $\mathcal B_{\text{min}} = \vec\beta_1$. In cases where $\vec\beta_1 < \sigma$ and $\mathcal B\so \geq \mathcal B_{\text{min}}$, the requirement can be strictly enforced by redefining the reward vector $r\st$ as 
    \begin{align*}
        r\st 
        \!=\! \begin{cases}
        q_{\text{min}} \vec\beta + \bar r + \tilde c(I\st) - \tilde c(\bar I) & \text{if $\mathcal B\st \!=\! \mathcal B_{\text{min}}$, \!$q\st \!\leq\! q_{\text{min}}$} \\
        q\st \vec\beta + \bar r + \tilde c(I\st) - \tilde c(\bar I) & \text{otherwise},
        \end{cases}
    \end{align*}
    where $q_{\text{min}}$ is selected to satisfy $\mathcal B_{\text{min}} = \vec\beta' C(q_{\text{min}} \vec\beta + \bar r - \tilde c (\bar I)) < \bar\beta$, and $\mathcal B_{\text{min}}$ can be chosen arbitrarily close to $\sigma$.
\end{remark}
}


\change{
In light of Theorem~\ref{theorem:stability}, the budget constraint (P2) can be expressed as $(\bar q \vec\beta + \bar r)'C(\bar q \vec\beta + \bar r - \tilde c (\bar I)) \leq c^\ast$ where $c^\ast$ is the cost budget, and $\bar q$ is the limit of $q\st$ for given $\bar r$ and $\bar \beta$. Consider the optimal choice $\bar r = r^\ast$ and $\bar \beta = \beta^\ast$, where $r^\ast$ and $\beta^\ast$ are part of the optimal solution to \eqref{eq:betadef}. In this case, we can conclude that $\bar q = 0$ and thus, ${r^\ast}'C(r^\ast - \tilde c (I^\ast))$ represents the average cost that the social planner needs to spend to maintain the optimal transmission rate $\beta^\ast$ when the closed-loop model reaches equilibrium. As a corollary of Theorem~\ref{theorem:stability}, we can show that with $\bar r = r^\ast$ and $\bar \beta = \beta^\ast$, the reward vector $r\st$ converges to $r^\ast$, and the transmission rate $\mathcal B\st$ converges to the minimum $\beta^\ast$ that satisfies the budget constraint.
}

\begin{corollary} \label{corollary:stability}
    Suppose $\bar r$ and $\bar \beta$ are determined as the optimal solution to \eqref{eq:betadef}: $\bar r = r^\ast$ and $\bar \beta = \beta^\ast = \vec\beta' C(r^\ast - \tilde c (I^\ast))$. For the feedback interconnection of (EPG) and (PBR EDM), assume that $\mathcal B\st \geq \mathcal B_{\text{min}}, ~\forall t \geq 0$, where $\beta^\ast > \mathcal B_{\text{min}} > \sigma$. Then, the following hold:
    \begin{enumerate}
        \item $\lim_{t \to \infty} \mathcal B\st = \beta^\ast$, and

        \item $\lim_{t \to \infty} r\st = r^\ast$ while satisfying ${r^\ast}' C(r^\ast - \tilde c (I^\ast)) \leq c^\ast$.
    \end{enumerate}
\end{corollary}

\change{
\begin{remark}
The following are remarks regarding Theorem~\ref{theorem:stability} and Corollary~\ref{corollary:stability}.
\begin{enumerate}
\item For the planner to compute $r^\ast$ and $\beta^\ast$ using \eqref{eq:betadef}, the knowledge of the choice function $C$ is essential. In \S\ref{sec:choice_function_learning}, we consider a scenario where $C$ is initially unknown to the planner. In such cases, $C$ must be estimated based on data gathered over time from the agents' strategy selections.

\item The proof of Theorem~\ref{theorem:stability} adopts the Lyapunov stability technique which, in conjunction with Corollary~\ref{corollary:stability}, implies that the infectious population $I\st$ converges to $I^\ast$. In other words, the reward mechanism defined by (EPGc,d) asymptotically attains the minimum achievable infectious population while asymptotically satisfying the given budget constraint.

\item The higher the value of $\kappa$, in \eqref{eq:G}, the faster the dynamics of $q\st$. Consequently, with large $\kappa$, the planner can more quickly adjust $q\st$ to control the spread of the epidemic. To see this, consider the function $\mathscr S (\mathcal I, \mathcal R, \mathcal B)$, defined in \eqref{eq:storage_function_for_EPG} and used to establish the convergence results in Theorem~\ref{theorem:stability}. Here, $\mathcal I$ and $\mathcal R$ are given as $\mathcal I = \mathcal B I$ and $\mathcal R = \mathcal B R$, respectively. 
The following conditions hold: 
\begin{align*}
    &\mathscr S (\mathcal I \st, \mathcal R \st, \mathcal B \st) = 0 \implies \dot I \st = 0 \\
    &\frac{\mathrm d \mathscr S}{\mathrm d t} (\mathcal I\st, \mathcal R\st, \mathcal B\st) \!=\! -\kappa^{-1} \! G (I\st, R\st, x\st, q\st) \dot{\mathcal B}\st \\
    & \qquad \qquad \qquad -(\mathcal I\st - \hat{\mathcal I}\st)^2 - \frac{\omega}{\gamma}(\mathcal R\st - \hat{\mathcal R}\st)^2
\end{align*}

We explain how large $\kappa$ facilitates the reduction of $\mathscr S (\mathcal I\st, \mathcal R\st, \mathcal B\st)$, thereby slowing the growth of the infectious population $I\st$, particularly when $\dot I\st > 0$. To achieve this, the planner can incentivize agents to adopt strategies that minimize the upper bound $-\kappa^{-1} G (I\st, R\st, x\st, q\st) \dot{\mathcal B}\st$ of $\frac{\mathrm d \mathscr S}{\mathrm d t} (\mathcal I\st, \mathcal R\st, \mathcal B\st)$, which directly depends on $q\st$.
When $\kappa$ is sufficiently large, $\kappa^{-1} G (I\st, R\st, x\st, q\st) < 0$ leads to a rapid decrease in $q\st$, which in turn reduces $\dot{\mathcal B}\st = \vec\beta' \argmax_{z \in \mathrm{int} (\mathbb X)} ( z' p \st - Q (z) ) - \mathcal B \st $ by driving  $\vec\beta' \argmax_{z \in \mathrm{int} (\mathbb X)} ( z' p \st - Q (z) )$ towards its lower bound $\vec\beta_1$, as indicated by Lemma~\ref{lemma:c_increasing_function} in Appendix~\ref{sec:proof_stability_theorem}. Conversely, when $\kappa^{-1} G (I\st, R\st, x\st, q\st) > 0$, the variable $q \st$ increases rapidly, leading to an increase in $\dot{\mathcal B}\st$ as $\vec\beta' \argmax_{z \in \mathrm{int} (\mathbb X)} ( z' p \st - Q (z) )$ moves towards its upper bound $\vec\beta_n$.
\end{enumerate}
\end{remark}
}

\begin{figure}
    \centering
    \begin{subfigure}{0.24\textwidth}
        \centering
        \ifnum \useTIKZ>0
            \resizebox{\textwidth}{!}{\input{figures/kappa_logit.b.SIRS_EDM_I_ratio_0.15_nu3.0.tikz}}
        \else
            \includegraphics[width=\textwidth,trim={15mm 15mm 10mm 15mm},clip]{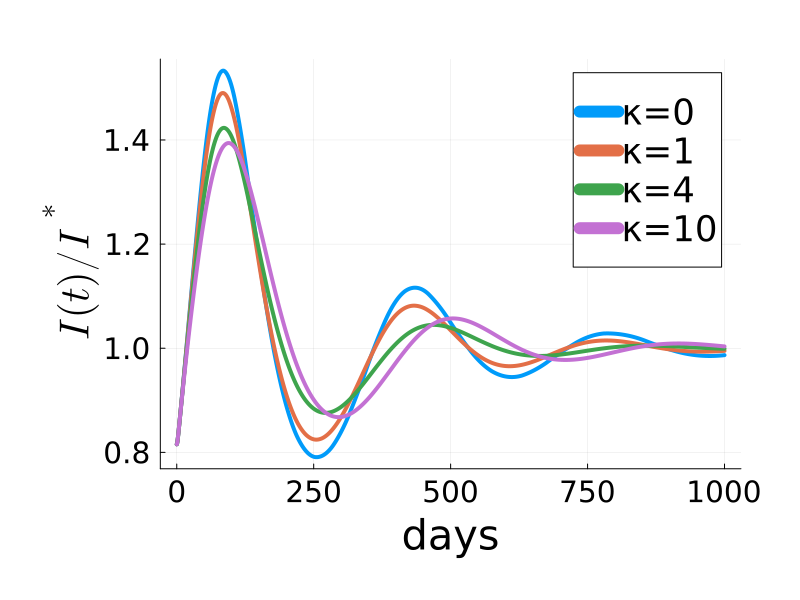}
        \fi
        \caption{}
        \label{fig:exampleKappaLogit:a}
    \end{subfigure}%
    \begin{subfigure}{0.24\textwidth}
        \centering
        \ifnum \useTIKZ>0
            \resizebox{\textwidth}{!}{\input{figures/kappa_logit.a.SIRS_EDM_cost_0.15_nu3.0.tikz}}
        \else
            \includegraphics[width=\textwidth,trim={15mm 15mm 10mm 15mm},clip]{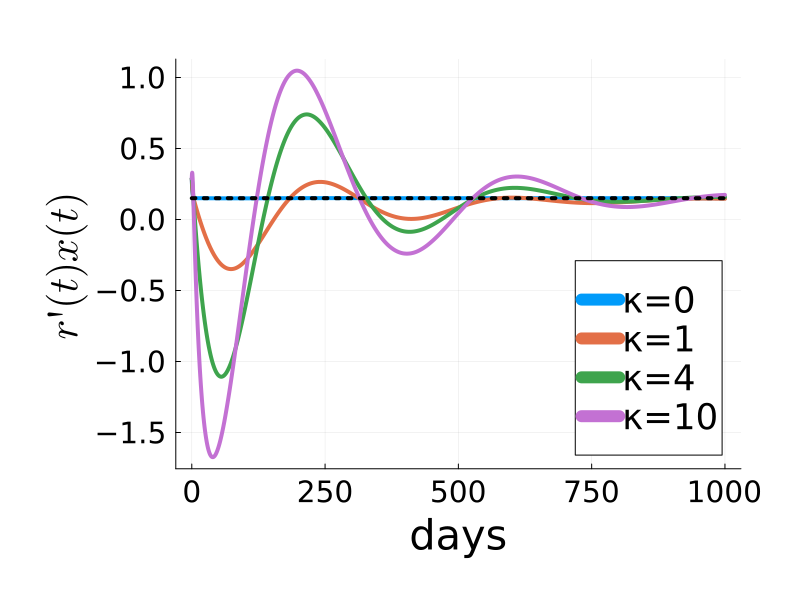}
        \fi
        \caption{}
        \label{fig:exampleKappaLogit:b}
    \end{subfigure}
    \caption{Simulation results for Example~\ref{ex:exampleLogit} illustrating (a) the infectious fraction $I\st$ of the population with respect to $I^\ast$ and (b) the average cost $r'\st x\st$, evaluated with different values of $\kappa$ for \eqref{eq:G}. 
    }
    \label{fig:exampleKappaLogit}
\end{figure}

The following example illustrates the convergence result of Corollary~\ref{corollary:stability}.
\begin{example} \label{ex:exampleLogit}
    Consider that (PBR EDM) is defined by the logit learning rule with $\mu=1$ and is interconnected with (EPG), where $\gamma=0.1$ (infectiousness period $\sim$ 10 days), $\sigma=\gamma$, $\omega=0.005$ (immunity period $\sim$ 200 days), and $\upsilon=3$. With the transmission rates $\vec{\beta} = (0.15,0.19)$, intrinsic cost 
    $c (I) = (0.2038 - 0.2 I, 0)$,
    and budget $c^*=0.15$, we obtain $r^* \approx (0.287, 0)$ and $\beta^* = 0.1691$ from \eqref{eq:betadef} under which $x\st$ and $(I\st, R\st)$ converge to $x^* = C(r^\ast - \tilde c{(I^*)}) \approx (0.522, 0.478)$ and $(I^*,R^*) \approx (0.019, 0.389)$, respectively. 

    Using the initial condition $x\so = (1,0)$, $I\sta{0} = 0.0158$, $R\sta{0} = 0.3170$, and $q\so = 0$, 
    Fig.~\ref{fig:exampleKappaLogit} illustrates the ratio $I\st/I^\ast$ and the instantaneous average cost $r'\st x\st$ with varying values of $\kappa$ in (EPGc). We can observe that $I\st/I^\ast$ converges to $1$ in Fig. \ref{fig:exampleKappaLogit:a}, indicating that the infectious population $I\st$ approaches the endemic equilibrium $I^\ast$, and $r'\st x\st$ converges to $c^\ast$, marked by the dotted horizontal line in Fig.~\ref{fig:exampleKappaLogit:b}. In addition, as we can observe from the figure, increasing $\kappa$ will reduce the overshoot in $I\st/I^\ast$ (see Fig.~\ref{fig:exampleKappaLogit:a}) at the expense of larger instantaneous cost (see Fig.~\ref{fig:exampleKappaLogit:b}).
\end{example}



\subsection{Anytime Bound} \label{sec:anytime_bounds}
\change{
We establish an anytime bound on the infectious fraction $I\st$ of the population.
By following the same arguments as in \cite[\S4]{MARTINS2023111016}, we proceed with defining
\begin{align} \label{eq:upper_bound_on_infectious_fraction}
    \pi_\upsilon (\alpha_\kappa) = \bar I^{-1} \sup \left\{ I \,|\, \mathscr S (\mathcal I, \mathcal R, \mathcal B) \leq \alpha_\kappa \right\},
\end{align}
where $\mathcal I = \mathcal BI$, $\mathcal R = \mathcal BR$, $\bar I = \eta (1 - \sigma / \bar \beta)$ for fixed $\bar \beta \in (\sigma, \vec\beta_n)$, and $\mathscr S(\mathcal I, \mathcal R, \mathcal B)$ is a function defined in \eqref{eq:storage_function_for_EPG}. The value $\pi_\upsilon (\alpha_\kappa)$ represents the upper bound of $I\st/\bar I$, where $\pi_\upsilon\so = 1$ and $\alpha_\kappa$ is determined by the initial condition $(I\so, R\so, q\so, x\so)$ of the closed-loop model. Specifically, $\alpha_\kappa$ is given by $\alpha_\kappa = \mathcal \kappa^{-1} \mathcal S(x\so, p\so) + \mathscr S(\mathcal I\so, \mathcal R\so, \mathcal B\so)$, where $\mathcal S(x, p)$ is the $\delta$-storage function of (PBR EDM) defined in \eqref{eq:delta_storage_function}.

To be more specific, according to \eqref{eq:lyapunov_function_time_derivative}, the function $\kappa^{-1} \mathcal S(x\st, p\st) + \mathscr S(\mathcal I\st, \mathcal R\st, \mathcal B\st)$ is decreasing in $t$ and, hence, it holds that
\begin{align} \label{eq:bound_on_epg_storage_function}
    \mathscr S(\mathcal I\st, \mathcal R\st, \mathcal B\st)
    \leq \alpha_\kappa, ~ \forall t \geq 0.
\end{align}
From \eqref{eq:upper_bound_on_infectious_fraction}, we can establish that $I\st \leq \bar I \pi_\upsilon (\alpha_\kappa)$ holds for all $t \geq 0$. Additionally, note that the constant $\alpha_\kappa$ decreases as $\kappa$ increases, as does the upper bound $\bar I \pi_\upsilon (\alpha_\kappa)$, as observed in Fig.~\ref{fig:exampleKappaLogit:a}.
}

\change{
Now, to evaluate how the choice of the parameters $\bar r$ and $\bar\beta$ affects the anytime bound, consider that the state of the closed-loop model, consisting of (EPG) and (PBR EDM), starts from the endemic equilibrium resulting from a prior use of (EPG) with $r^o$ and $\beta^o$, respectively, as the stationary reward vector and endemic transmission rate. Hence, $I\so = \hat I\so = \eta ( 1 - \sigma / \mathcal B\so )$ and $R\so = \hat R\so = (1 - \eta) ( 1 - \sigma / \mathcal B\so )$ hold.
Suppose the social planner adopts $\bar r$ and $\bar \beta$ as the revised stationary reward vector and target endemic transmission rate, respectively.
Note that in this case, $\mathcal B\so = \beta^o$, $p\so = q\so \vec\beta + \bar r - \tilde c (\bar I)$, and $\mathscr S (\mathcal I\so, \mathcal R\so, \mathcal B\so) = \upsilon^2 \tilde \beta^2 / 2$ hold, where $\bar I = \eta (1 - \sigma/\bar \beta)$ and $\tilde \beta = \beta^o - \bar \beta$.
Also, we can evaluate the $\delta$-storage function $\mathcal S$ of (PBR EDM) at $t=0$ as
\begin{align}
    \mathcal S(x\so, p\so)
     & = \max_{z \in \mathrm{int}(\mathbb X)} ( z' (q\so\vec\beta + \bar r - \tilde c (\bar I)) - Q(z) ) \nonumber     \\
     & \qquad - ( x'\so (q\so\vec\beta + \bar r - \tilde c (\bar I)) - Q(x\so) ) \nonumber                             \\
     & = \max_{z \in \mathrm{int}(\mathbb X)} ( z' (q\so\vec\beta + \bar r - \tilde c (\bar I)) - Q(z) ) \nonumber     \\
     & \qquad - \max_{z \in \mathrm{int}(\mathbb X)} ( z' (q\so\vec\beta + r^o - \tilde c (I^o)) - Q(z) ) \nonumber \\
     & \qquad - x'\so \left( \bar r - \tilde c(\bar I) - (r^o - \tilde c (I^o)) \right),
\end{align}
where $I^o = \eta (1 - \sigma/\beta^o)$.
To derive the latter equality, we use the fact that before the revision to $\bar r$ and $\bar \beta$, the state was at the previous endemic equilibrium. Consequently, $\mathcal S(x\so, q\so\vec\beta + r^o - \tilde c (I^o)) = 0$ holds. With fixed $r^o$ and $\beta^o$, we can express $\mathcal S(x\so, p\so)$ as a continuous function $B_{\mathcal S}(\tilde r, \tilde \beta)$ of $\tilde r$ and $\tilde \beta$ satisfying $B_{\mathcal S} (0,0) = 0$, where $\tilde r = r^o - \bar r$.\footnote{\change{The function $B_{\mathcal S}$ is continuous since $\max_{z \in \mathrm{int}(\mathbb X)} ( z' (q\so\vec\beta + r) - Q(z) )$ is differentiable in $r$ \cite{Hofbauer2002On-the-global-c} and $\bar r - \tilde c(\bar I) - (r^o - \tilde c (I^o))$ is continuous in $\tilde r$ and $\tilde \beta$.}}

In conjunction with \eqref{eq:upper_bound_on_infectious_fraction} and \eqref{eq:bound_on_epg_storage_function}, by selecting $\alpha_\kappa = \kappa^{-1} \mathcal S(x\so, p\so) + \mathscr S (\mathcal I\so, \mathcal R\so, \mathcal B\so) = \kappa^{-1} B_{\mathcal S}(\tilde r, \tilde \beta) + \upsilon^2 \tilde \beta^2 /2$, we can establish the anytime bound on $I\st$ as
\begin{align} \label{eq:anytime_bound_on_I_t}
    I\st \leq \bar I \pi_\upsilon (\kappa^{-1} B_{\mathcal S}(\tilde r, \tilde \beta) + \upsilon^2 \tilde \beta^2 /2), \, \forall t \geq 0.
\end{align}
Therefore, the bound \eqref{eq:anytime_bound_on_I_t} indicates how much the infectious fraction $I\st$ exceeds its target value $\bar I = \eta (1 - \sigma / \bar \beta)$ when the planner revises the stationary reward vector and endemic transmission rate to $\bar r$ and $\bar \beta$, respectively. Additionally, \eqref{eq:anytime_bound_on_I_t} highlights how the choice of $\kappa$ and $\upsilon$ influences the overshoot of $I\st$.

Using a similar analysis, we can establish an anytime bound on the instantaneous average reward $r’\st x\st$ as follows:
\begin{align*}
\sup \big\{ (p + \tilde c (I))' x \,\big|\, \kappa^{-1} \mathcal S (x,p) + \mathscr S (\mathcal I, \mathcal R, \mathcal B) \leq \alpha_\kappa \big\},
\end{align*}
where $p = q \vec\beta + \bar r - \tilde c(\bar I)$, $\mathcal B = \vec\beta’ x$, $\mathcal I = \mathcal B I$, $\mathcal R = \mathcal B R$, $\bar I = \eta ( 1 - \sigma/\bar\beta)$, and $\alpha_\kappa = \kappa^{-1} \mathcal S(x\so, p\so) + \mathscr S(\mathcal I\so, \mathcal R\so, \mathcal B\so)$. 
}

\subsection{Learning One-Parameter Choice Function $C^\mu$} \label{sec:choice_function_learning}
In \S\ref{sec:convergence_results} and \S\ref{sec:anytime_bounds}, we discussed computing the optimal stationary reward vector $r^\ast$ and endemic transmission rate $\beta^\ast$ using the minimization \eqref{eq:betadef}, and establishing the anytime bound \eqref{eq:anytime_bound_on_I_t}. To solve the minimization and compute the bound, the planner needs to know the choice function $C$, which dictates how agents in the population revise their strategies. In this section, we discuss a scenario wherein the planner needs to estimate  $C$.

We focus on a class of one-parameter choice functions, motivated by empirical studies \cite{Rosenthal1989, GOEREE2005349} demonstrating the predictive power of such a class of models.\footnote{Also, as explained in Appendix~\ref{appendix_a}, the choice of probability distribution for the noise in \eqref{eq:perturbed_revision_protocol} does not substantially impact the outcomes of (EPG). This observation highlights that the one-parameter choice functions would be effective models in studying epidemic population games.}
To proceed, we express $C$ as
\begin{align} \label{eq:one-parameter_choice_function}
    C(p) = C^\mu(p) = \argmax_{z \in \mathrm{int} (\mathbb X)} ( z' p - \mu \bar Q (z) ),
\end{align}
where $\bar Q: \mathrm{int}(\mathbb X) \to \mathbb R$ is a known deterministic payoff perturbation whereas $\mu > 0$ is an unknown parameter that needs to be estimated. 
The parameter $\mu$ quantifies the level of the perturbation. To see this, according to \eqref{eq:perturbed_revision_protocol}, recall that the choice function is given as $C_i^\mu(p) = \mathbb P ( p_i + v_i \geq \max_{1 \leq j \leq n} (p_j + v_j ) )$, where $v_1, \cdots, v_n$ are random variables associated with the deterministic perturbation $\mu \bar Q$.
By defining
$C_i^1(p) = \mathbb P ( p_i + \bar v_i \geq \max_{1 \leq j \leq n} (p_j + \bar v_j ) )$, we can establish that $v_i = \mu \bar v_i, ~ \forall i \in \{1, \cdots, n\}$, where $\bar v_1, \cdots, \bar v_n$ are random variables associated with $\bar Q$.
Consequently, $\mu$ can be interpreted as the scaling factor for $\bar v_i$. Hence, the larger the parameter $\mu$ is, the greater the noise is in the agent decision making. The assumption that $\bar Q$ is known implies that the planner knows the joint probability density function of $\bar v_1, \cdots, \bar v_n$.

In the following lemma, we examine an important property of the choice function $C^\mu$ that is useful in the estimation of $\mu$.

\begin{lemma} \label{proposition:estimated_perturbation_parameter}
    Let $C^\mu$ be a one-parameter choice function \eqref{eq:one-parameter_choice_function}. For any given $r \in \mathbb R_{\geq 0}^n$ and $\tilde c (I) \in \mathbb R_{\geq 0}^n$ associated with fixed $I$, where not all entries of $r - \tilde c (I)$ are identical,
    $(r - \tilde c (I))' C^\mu (r-\tilde c (I))$ is a decreasing function of $\mu$.
\end{lemma}

The proof is provided in Appendix~\ref{sec:proof_proposition_estimated_perturbation_parameter}.
Consequently, the choice function $C^\mu$ satisfies $(r - \tilde c (I))' C^{\hat \mu} (r - \tilde c (I)) = (r - \tilde c (I))' C^{\mu}(r - \tilde c (I))$ if and only if $\hat \mu = \mu$ holds. Hence, when $(r, C^\mu (r - \tilde c (I)))$ is given as data for the estimation of $\mu$, the social planner can learn unique $\mu$. In what follows, we discuss how the planner can devise a parameter estimation scheme based on Lemma~\ref{proposition:estimated_perturbation_parameter} and then use it in conjunction with (EPG) to determine $r\st$ for the population.

\change{At the beginning of a pandemic, a social planner, who is initially unaware of the parameter $\mu$, can employ (EPG) with an initial selection of $\bar r$ and $\bar \beta$ to curb the spread of the pandemic. Simultaneously, the planner can collect data on the agents' strategy selections to estimate $\mu$. For data collection, we consider a survey method in which randomly selected agents are asked which strategies they would choose given a fixed reward vector $r$, which may differ from $\bar r$ used in (EPG). Note that each agent's response represents a specific strategy randomly drawn from the probability distribution $C^{\mu} (r - \tilde c(I))$. The following proposition outlines the method for calculating probabilistic upper and lower bounds of $\mu$ using the survey data.}

\change{
\begin{proposition} \label{prop:parameter_estimation}
    Let $r = \tilde r + \tilde c(I)$ represent the reward vector used in the survey, where $I$ denotes the current infectious population and $\tilde r$ is a vector that satisfies the condition:
\begin{align} \label{eq:condition_for_popoviciu_inequlity}
    \max_{1 \leq i \leq n} \tilde r_i - \min_{1 \leq i \leq n} \tilde r_i = 2.
\end{align}
Let $\{\xi_k\}_{k=1}^K \subset \{e_1, \cdots, e_n\}$ represent the strategy selections of agents based on the the probability distribution $C^{\mu}(r - \tilde c(I))$, where the standard basis vector $e_i$ corresponds to the selection of strategy~$i$. Given a positive constant $\epsilon$, compute $\mu_L$ and $\mu_U$ satisfying
\begin{subequations} \label{eq:bounds_on_mu}
\begin{align}
    \tilde r' C^{\mu_L} (\tilde r) &= \frac{1}{K} \sum_{k=1}^K \tilde r' \xi_k + \epsilon \\
    \tilde r ' C^{\mu_U} (\tilde r) &= \frac{1}{K} \sum_{k=1}^K \tilde r' \xi_k - \epsilon.
\end{align}
\end{subequations}
We can establish the following probabilistic bounds on $\mu$:
\begin{align} \label{eq:probabilistic_bounds_on_mu}
    \mathbb P \left( \mu_L \leq \mu \leq \mu_U \right) \geq 1 - 1/\epsilon^2 K.
\end{align}
\end{proposition}
\vspace{.5em}
Once enough data has been collected to estimate $\mu$ with the required accuracy, the planner can optimize the choice of $\bar r$ and $\bar \beta$ using \eqref{eq:betadef}. Specifically, with a sufficiently large number of data points $K$, a small enough $\epsilon$ can be chosen to minimize the gap between the lower bound $\mu_L$ and upper bound $\mu_U$, while maintaining a desired confidence level of $1 - 1/\epsilon^2 K$.
}


During this parameter learning phase, since the planner does not know the actual value of $\mu$, it can be difficult to predict the limit $\bar q$ of $q\st$ and ensure that the budget constraint $(\bar q \vec\beta + \bar r)' C^\mu(\bar q \vec\beta + \bar r - \tilde c (\bar I)) \leq c^\ast$ is satisfied. To address this issue, using the bounds on $\mu$, we explain how the planner can estimate an upper bound on $(\bar q \vec\beta + \bar r)' C^\mu(\bar q \vec\beta + \bar r - \tilde c (\bar I))$ and use the bound to meet the budget requirement.

Suppose $\mu$ belongs to $[\mu_L, \mu_U]$. Then, the upper bound on $(\bar q \vec\beta + \bar r)' C^\mu(\bar q \vec\beta + \bar r - \tilde c (\bar I))$ can be derived as
\begin{align} \label{eq:long_term_cost_estimate}
    \max_{\mu_L \leq \mu \leq \mu_U} (\bar q (\mu) \vec \beta + \bar r)' C^{\mu} (\bar q (\mu) \vec \beta + \bar r - \tilde c (\bar I)),
\end{align}
where $\bar q(\mu)$ is a solution, which depends on $\mu$, to $\bar \beta = \vec \beta' C^{\mu} (\bar q(\mu) \vec\beta + \bar r - \tilde c (\bar I))$.
Ideally, we want the evaluation of \eqref{eq:long_term_cost_estimate} to be computationally tractable so that the planner can assess the upper bound \eqref{eq:long_term_cost_estimate} over a wide range of $\bar r, \bar \beta$ and select the one that minimizes the endemic transmission rate $\bar \beta$ subject to the budget constraint.

In the following proposition, we derive technical conditions under which the upper bound \eqref{eq:long_term_cost_estimate} can be computed using a single run of Newton's method. For this purpose, given $\bar\beta$, we fix $\bar r = \tilde c (\bar I)$ with $\bar I = \eta (1 - \sigma / \bar\beta)$ and assume $\tilde c' (\bar I) C^{1}\so < c^\ast$.

\begin{figure*}
    \centering
    \begin{subfigure}[b]{0.23\textwidth}
        \centering
        \ifnum \useTIKZ>0
            \resizebox{\textwidth}{!}{\input{figures/cost_bound_mu_1.tikz}}
        \else
            \includegraphics[width=\textwidth,trim={15mm 13mm 6mm 15mm},clip]{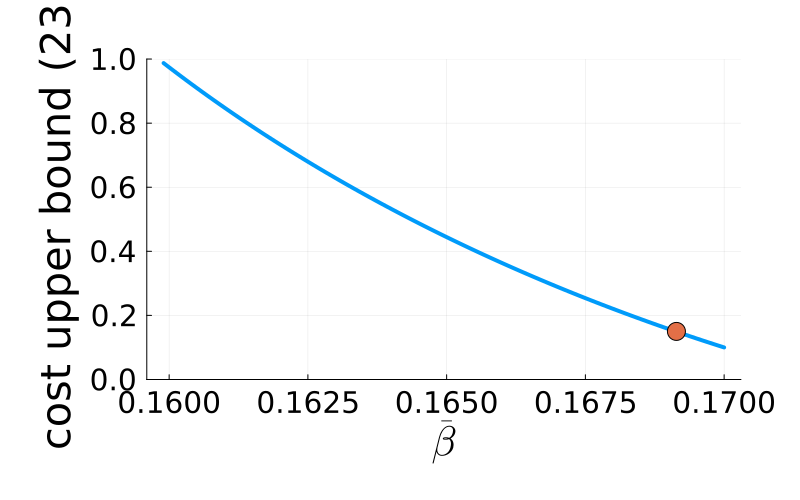}
        \fi
        \caption{$\mu_U = 1$}
        \label{fig:cost_upper_bound_a}
    \end{subfigure}%
    \begin{subfigure}[b]{0.22\textwidth}
        \centering
        \ifnum \useTIKZ>0
            \resizebox{\textwidth}{!}{\input{figures/cost_bound_mu_2.tikz}}
        \else
            \includegraphics[width=\textwidth,trim={15mm 13mm 6mm 15mm},clip]{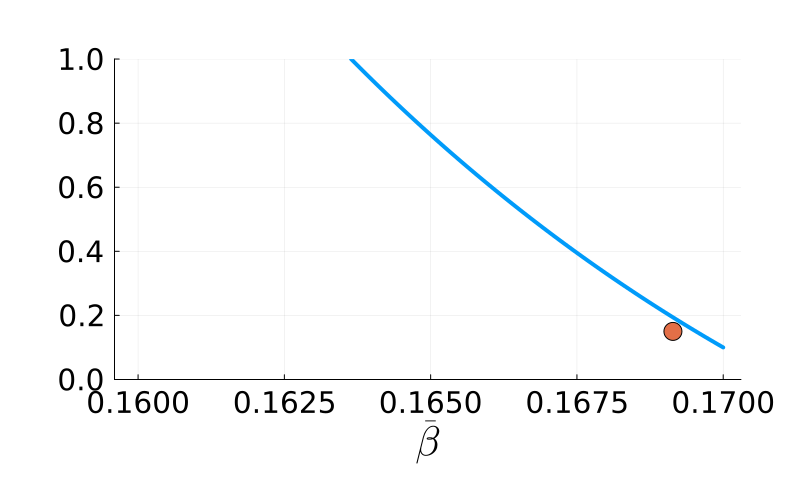}
        \fi
        \caption{$\mu_U = 2$}
        \label{fig:cost_upper_bound_b}
    \end{subfigure}%
    \begin{subfigure}[b]{0.22\textwidth}
        \centering
        \ifnum \useTIKZ>0
            \resizebox{\textwidth}{!}{\input{figures/cost_bound_mu_5.tikz}}
        \else
            \includegraphics[width=\textwidth,trim={15mm 13mm 6mm 15mm},clip]{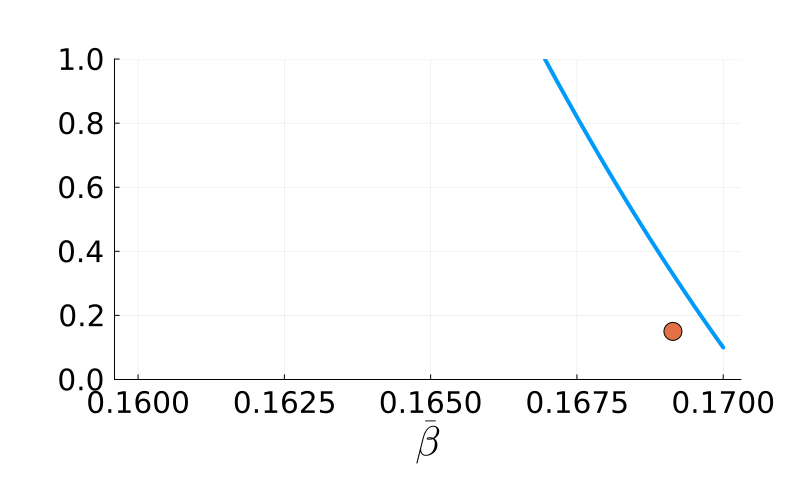}
        \fi
        \caption{$\mu_U = 5$}
        \label{fig:cost_upper_bound_c}
    \end{subfigure}%
    \begin{subfigure}[b]{0.265\textwidth}
        \ifnum \useTIKZ>0
            \resizebox{\textwidth}{!}{

\begin{tikzpicture}[/tikz/background rectangle/.style={fill={rgb,1:red,1.0;green,1.0;blue,1.0}, fill opacity={1.0}, draw opacity={1.0}}, show background rectangle]
\begin{axis}[point meta max={nan}, point meta min={nan}, legend cell align={left}, legend columns={1}, title={}, title style={at={{(0.5,1)}}, anchor={south}, font={{\fontsize{14 pt}{18.2 pt}\selectfont}}, color={rgb,1:red,0.0;green,0.0;blue,0.0}, draw opacity={1.0}, rotate={0.0}, align={center}}, legend style={color={rgb,1:red,0.0;green,0.0;blue,0.0}, draw opacity={1.0}, line width={1}, solid, fill={rgb,1:red,1.0;green,1.0;blue,1.0}, fill opacity={1.0}, text opacity={1.0}, font={{\fontsize{14 pt}{18.2 pt}\selectfont}}, text={rgb,1:red,0.0;green,0.0;blue,0.0}, cells={anchor={center}}, at={(1.02, 1)}, anchor={north west}}, axis background/.style={fill={rgb,1:red,1.0;green,1.0;blue,1.0}, opacity={1.0}}, anchor={north west}, xshift={6.0mm}, yshift={-6.0mm}, width={191.2mm}, height={115.0mm}, scaled x ticks={false}, xlabel={$\mu_U$}, x tick style={color={rgb,1:red,0.0;green,0.0;blue,0.0}, opacity={1.0}}, x tick label style={color={rgb,1:red,0.0;green,0.0;blue,0.0}, opacity={1.0}, rotate={0}}, xlabel style={at={(ticklabel cs:0.5)}, anchor=near ticklabel, at={{(ticklabel cs:0.5)}}, anchor={near ticklabel}, font={{\fontsize{14 pt}{18.2 pt}\selectfont}}, color={rgb,1:red,0.0;green,0.0;blue,0.0}, draw opacity={1.0}, rotate={0.0}}, xmajorgrids={true}, xmin={0.8829999999999996}, xmax={5.017000000000005}, xticklabels={{$1$,$2$,$3$,$4$,$5$}}, xtick={{1.0,2.0,3.0,4.0,5.0}}, xtick align={inside}, xticklabel style={font={{\fontsize{12 pt}{15.600000000000001 pt}\selectfont}}, color={rgb,1:red,0.0;green,0.0;blue,0.0}, draw opacity={1.0}, rotate={0.0}}, x grid style={color={rgb,1:red,0.0;green,0.0;blue,0.0}, draw opacity={0.1}, line width={0.5}, solid}, axis x line*={left}, x axis line style={color={rgb,1:red,0.0;green,0.0;blue,0.0}, draw opacity={1.0}, line width={1}, solid}, scaled y ticks={false}, ylabel={$\bar \beta_{\text{min}}$}, y tick style={color={rgb,1:red,0.0;green,0.0;blue,0.0}, opacity={1.0}}, y tick label style={color={rgb,1:red,0.0;green,0.0;blue,0.0}, opacity={1.0}, rotate={0}}, ylabel style={at={(ticklabel cs:0.5)}, anchor=near ticklabel, at={{(ticklabel cs:0.5)}}, anchor={near ticklabel}, font={{\fontsize{14 pt}{18.2 pt}\selectfont}}, color={rgb,1:red,0.0;green,0.0;blue,0.0}, draw opacity={1.0}, rotate={0.0}}, ymajorgrids={true}, ymin={0.16910560913085934}, ymax={0.16982180786132808}, yticklabels={{$0.1692$,$0.1693$,$0.1694$,$0.1695$,$0.1696$,$0.1697$,$0.1698$}}, ytick={{0.16920000000000002,0.1693,0.1694,0.1695,0.1696,0.16970000000000002,0.1698}}, ytick align={inside}, yticklabel style={font={{\fontsize{12 pt}{15.600000000000001 pt}\selectfont}}, color={rgb,1:red,0.0;green,0.0;blue,0.0}, draw opacity={1.0}, rotate={0.0}}, y grid style={color={rgb,1:red,0.0;green,0.0;blue,0.0}, draw opacity={0.1}, line width={0.5}, solid}, axis y line*={left}, y axis line style={color={rgb,1:red,0.0;green,0.0;blue,0.0}, draw opacity={1.0}, line width={1}, solid}, colorbar={false}]
    \addplot[color={rgb,1:red,0.0;green,0.6056;blue,0.9787}, name path={64}, draw opacity={1.0}, line width={4}, solid]
        table[row sep={\\}]
        {
            \\
            1.0  0.16912587890624997  \\
            1.1  0.16919667968749996  \\
            1.2000000000000002  0.16925527343749996  \\
            1.3000000000000003  0.16930654296874997  \\
            1.4000000000000004  0.1693529296875  \\
            1.5000000000000004  0.1693919921875  \\
            1.6000000000000005  0.16942617187499998  \\
            1.7000000000000006  0.16945791015625  \\
            1.8000000000000007  0.16948598632812498  \\
            1.9000000000000008  0.16951162109374995  \\
            2.000000000000001  0.16953359375  \\
            2.100000000000001  0.16955434570312497  \\
            2.200000000000001  0.16957387695312498  \\
            2.300000000000001  0.169590966796875  \\
            2.4000000000000012  0.16960683593749998  \\
            2.5000000000000013  0.16962148437499996  \\
            2.6000000000000014  0.1696361328125  \\
            2.7000000000000015  0.16964833984375  \\
            2.8000000000000016  0.16966054687499998  \\
            2.9000000000000017  0.16967153320312497  \\
            3.0000000000000018  0.16968251953124996  \\
            3.100000000000002  0.16969228515625  \\
            3.200000000000002  0.169700830078125  \\
            3.300000000000002  0.169709375  \\
            3.400000000000002  0.169717919921875  \\
            3.500000000000002  0.16972524414062498  \\
            3.6000000000000023  0.16973256835937497  \\
            3.7000000000000024  0.16973989257812497  \\
            3.8000000000000025  0.16974599609374996  \\
            3.9000000000000026  0.1697527099609375  \\
            4.000000000000003  0.16975881347656252  \\
            4.100000000000003  0.169764306640625  \\
            4.200000000000003  0.16976979980468748  \\
            4.3000000000000025  0.16977529296875  \\
            4.400000000000003  0.16978017578125  \\
            4.5000000000000036  0.16978505859374998  \\
            4.600000000000003  0.16978933105468746  \\
            4.700000000000003  0.16979360351562497  \\
            4.800000000000003  0.16979787597656248  \\
            4.900000000000004  0.16980153808593745  \\
        }
        ;
\end{axis}
\end{tikzpicture}}
        \else
            \includegraphics[width=\textwidth,trim={15mm 15mm 10mm 15mm},clip]{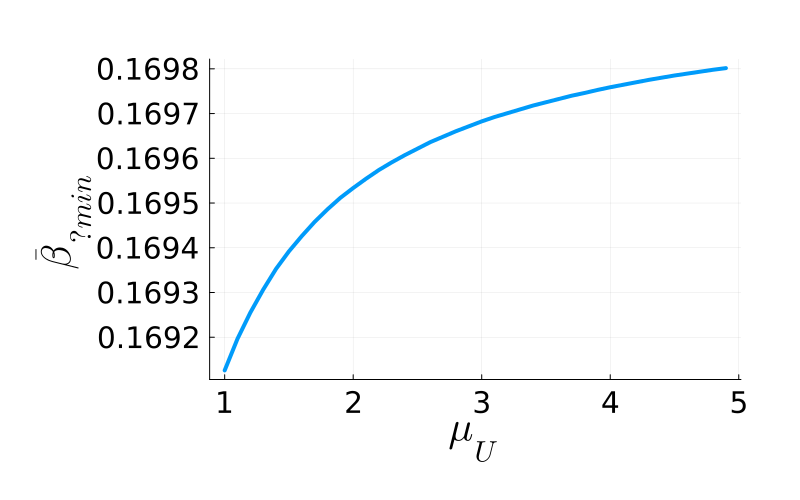}
        \fi
        \caption{}
        \label{fig:cost_upper_bound_d}
    \end{subfigure}%
    \caption{Plots (a)--(c) illustrate how the cost upper bound \eqref{eq:cost_upper_bound} varies depending on $\bar \beta$ for each fixed $\mu_U$: (a) $\mu_U=1$, (b) $\mu_U=2$, and (c) $\mu_U=5$. The red circle indicates the optimal endemic transmission rate $\beta^\ast =0.1691$ and the budget $c^\ast = 0.15$. Plot (d) depicts the smallest $\bar \beta_{\text{min}}$ under which the cost upper bound does not exceed the budget $c^\ast = 0.15$ for $\mu_U \in [1, 5]$, i.e., $\bar \beta_{\text{min}} = \min \{\bar \beta > 0 \,\big|\, \mu_U \lambda (\bar \beta \!-\! \vec\beta_n) + \tilde c' (\bar I) \argmax_{z \in \mathrm{int} (\mathbb X)} ( z' (\lambda \vec\beta) \!-\! \bar Q (z) ) \leq c^\ast \}$.}
    \label{fig:cost_upper_bound}
\end{figure*}

\begin{proposition} \label{proposition:cost_upper_bound}
    Suppose $\mu$ lies within the interval $[\mu_L, \mu_U]$ and $\bar \beta$ satisfies $\bar \beta < \vec \beta' C^{1}\so$.
    For fixed $\bar \beta$, the average stationary cost $(\bar q \vec\beta + \bar r)' C^\mu(\bar q \vec\beta + \bar r - \tilde c (\bar I))$ is upper bounded by
    \begin{align} \label{eq:cost_upper_bound}
        \mu_U \lambda (\bar \beta \!-\! \vec\beta_n) + \tilde c' (\bar I) \argmax_{z \in \mathrm{int} (\mathbb X)} ( z' (\lambda \vec\beta) \!-\! \bar Q (z) ).
    \end{align}
    The parameter $\lambda$ is a negative real number satisfying \\ $\bar \beta = \vec\beta' \argmax_{z \in \mathrm{int} (\mathbb X)} ( z' (\lambda \vec\beta) - \bar Q (z) )$, where $$\vec\beta' \argmax_{z \in \mathrm{int} (\mathbb X)} ( z' (\lambda \vec\beta) - \bar Q (z) )$$ is an increasing function of $\lambda$.
\end{proposition}

The proof is provided in Appendix~\ref{sec:proof_proposition_cost_upper_bound}.
By Proposition~\ref{proposition:cost_upper_bound}, the unique solution $\lambda$ in \eqref{eq:cost_upper_bound} can be found using Newton's method. Based on the results stated in the proposition, the planner can compute an appropriate $\bar \beta$ as follows: Starting from $\bar \beta = \vec\beta' C^{1}\so$, which incurs the average stationary cost of $\tilde c'(\bar I) C^{1}\so < c^\ast$, the planner can repeatedly assess \eqref{eq:cost_upper_bound} as it decreases the value of $\bar \beta$ until it finds the lowest $\bar \beta$ whose associated bound \eqref{eq:cost_upper_bound} satisfies the budget constraint.

Fig.~\ref{fig:cost_upper_bound} illustrates how the upper bound \eqref{eq:cost_upper_bound} varies depending on the values of $\mu_U$ and $\bar \beta$. The choice function is assumed to be logit with parameter $\mu=1$, and the parameters $\vec \beta$, $\tilde c$, and $c^\ast$ are given by $\vec \beta = (0.15, 0.19)$, $\tilde c (I) = (0.2038 - 0.2 I, 0)$, and $c^\ast = 0.15$, respectively. From the plots in Fig.~\ref{fig:cost_upper_bound}, we observe that as the bound on $\mu$ becomes loose, so does the estimated upper bound of the average stationary cost for the given budget. In particular, we can see from Fig.~\ref{fig:cost_upper_bound_a} that if the upper bound of $\mu$ is tight, i.e., $\mu_U = \mu$, then so is the bound \eqref{eq:cost_upper_bound} for the given $c^\ast = 0.15$.

\section{Simulations}
\label{sec:sims}

We now present simulations that illustrate our main results. In particular, we use a deployment scenario, detailed below, to demonstrate how the parameter learning approach for the choice function, along with the upper bound of the average stationary cost (as discussed in \S\ref{sec:choice_function_learning}), can be applied by the planner to infer the parameter $\mu$ of the choice function and to determine appropriate $\bar r, \bar \beta$ for (EPG).

\begin{enumerate}
    \item At the beginning of a pandemic, given a budget of $c^\ast$ and a known upper bound of $\mu$, the planner computes the minimum $\bar \beta$ such that the corresponding cost upper bound \eqref{eq:cost_upper_bound} satisfies the budget constraint with $\bar r = \tilde c (\bar I)$, where $\bar I = \eta (1 - \sigma / \bar\beta )$.

    \item By surveying randomly selected agents on their strategy choices with the reward vector $r = \tilde r + \tilde c (I)$, where $\tilde r$ is a constant vector and $I$ denotes the current infectious population, the planner computes the sample mean of the survey outcomes. Using Proposition~\ref{prop:parameter_estimation}, it then estimates the upper and lower bounds of $\mu$ for a given confidence level.

    \item After estimating the value of $\mu$ satisfying any required accuracy and confidence level, the planner computes the optimal solution $r^\ast, \beta^\ast$ for \eqref{eq:betadef} and revises $\bar r, \bar \beta$ accordingly. Using the revised $\bar r, \bar \beta$ and estimated $\mu$, the planner can establish the anytime bound on the infectious population $I\st$ based on \eqref{eq:upper_bound_on_infectious_fraction}.
    
\end{enumerate}

Consider that the agents adopt the logit learning rule with $\mu=1$ and the parameters of (EPG) are given by $\sigma = \gamma=0.1$,  $\omega=0.005$, $\kappa = 1$, and $\upsilon=3$.
With the transmission rates $\vec{\beta} = (0.15,0.19)$, intrinsic cost $c (I) = (0.2038 - 0.2 I, 0)$, and budget $c^*=1$, we obtain $r^* \approx (1.3248, 0)$ and $\beta^* \approx 0.1598$ from \eqref{eq:betadef} under which $I\st$ converges to $I^* \approx 0.0178$.
To define initial conditions for (EPG) and (PBR EDM), we assume that the majority of the agents use the costlier strategy at $t=0$, i.e., $x\so = (0.997, 0.003)$, and $I\sta{0} = 0.0159$, $R\sta{0} = 0.318$, and $q\so = 0$.

Suppose that the planner is aware of the range $(0, 5]$ for the parameter $\mu$, but not its exact value. By following the method explained in Proposition~\ref{proposition:cost_upper_bound}, the planner selects $\bar r = \tilde c (\bar I)$ and $\bar \beta \approx 0.167$ under which the upper bound \eqref{eq:cost_upper_bound} satisfies the budget constraint. The planner utilizes \eqref{eq:probabilistic_bounds_on_mu} with $\epsilon = 0.05$  to establish the lower and upper bounds of $\mu$ and to determine the number of data points required to achieve a confidence level of 0.95 as it gathers poll data from 1,000 randomly selected agents every 30 days.
Let $t_0 = 240$ be the day on which the estimated $\mu$ satisfies the estimation accuracy requirement. Once it does, the planner computes the estimate of $\mu$ based on the collected poll data, and the parameters $\bar r$, $\bar \beta$ are updated using the solution $r^\ast, \beta^\ast$ to \eqref{eq:betadef}. We consider the following two scenarios for the update of $\bar r$, $\bar \beta$.

\begin{figure}
    \centering
    \begin{subfigure}[b]{0.22\textwidth}
        \centering
        \ifnum \useTIKZ>0
            \resizebox{\textwidth}{!}{\input{figures/infectious.tikz}}
        \else
            \includegraphics[width=\textwidth,trim={15mm 13mm 12mm 15mm},clip]{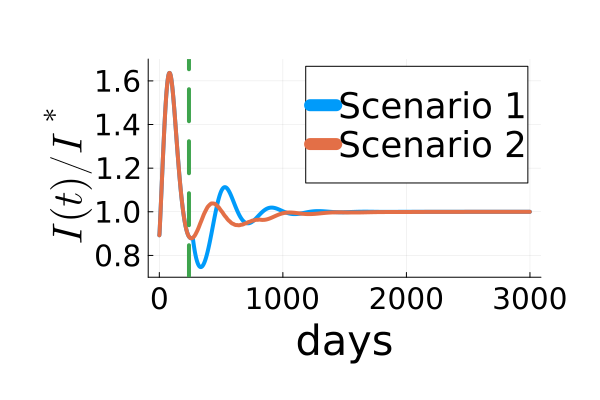}
        \fi
        \caption{}
        \label{fig:scenario_a}
    \end{subfigure}
    \begin{subfigure}[b]{0.22\textwidth}
        \centering
        \ifnum \useTIKZ>0
            \resizebox{\textwidth}{!}{\input{figures/average_reward.tikz}}
        \else
            \includegraphics[width=\textwidth,trim={15mm 13mm 12mm 15mm},clip]{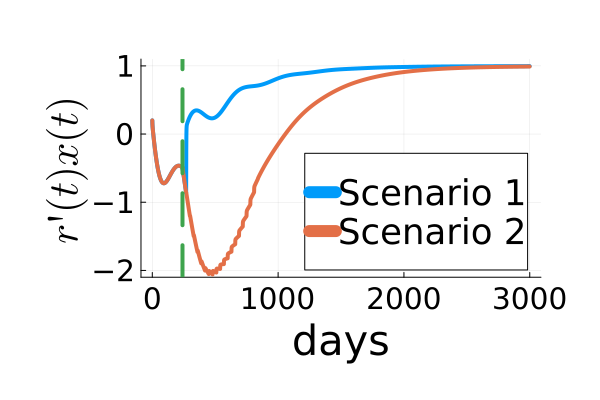}
        \fi
        \caption{}
        \label{fig:scenario_b}
    \end{subfigure}
    \caption{Simulation results for Scenario~1 and Scenario~2 illustrating (a) the infectious fraction $I\st$ of the population with respect to $I^\ast$ and (b) the average cost $r'\st x\st$, where the parameters $\bar r, \bar \beta$ of (EPG) are updated using the optimal $r^\ast, \beta^\ast$ beginning from the day $t_0 = 240$, as indicated by the green dotted vertical line.
    }
    \label{fig:scenario}
\end{figure}

In Scenario~1, the planner assigns $\bar r = r^\ast$ and $\bar \beta = \beta^\ast$ on the day $t_0$, whereas in Scenario~2, the planner selects $\bar \beta = \beta^\ast$ on the day $t_0$ and gradually changes $\bar r$ toward $r^\ast$. To implement the second scenario, after the day $t_0$, the planner evaluates
\begin{multline*}
    \alpha = \mathcal S(x\st, p\st) + \mathscr S(\mathcal I\st, \mathcal R\st, \mathcal B\st), \\ t = t_0+30, t_0+60, \cdots
\end{multline*}
at every $30$ days and updates $\bar r$ to the value that is closest to $r^\ast$ while ensuring that $\alpha$ remains below $0.0004$. According to \eqref{eq:upper_bound_on_infectious_fraction}, this corresponds to a $15\%$ overshoot in $I\st/I^\ast$. Fig.~\ref{fig:scenario} illustrates the simulation results for the two scenarios. As depicted in Fig.~\ref{fig:scenario_a}, the overshoot in the ratio $I\st/I^\ast$ is smaller when the parameters $\bar r, \bar \beta$ are judiciously tuned after $t_0$ using the anytime bound \eqref{eq:upper_bound_on_infectious_fraction}. 
But, this reduction in the second overshoot, which appears after $t_0$, 
requires penalizing the population (i.e., negative rewards) as can be observed in Fig.~\ref{fig:scenario_b}. 
Thus, maintaining non-negative rewards would demand increasing the rewards according to \eqref{eq:H_nonnegative_incentive}. 

\section{Conclusions and Future Research Plans}
\label{sec:conclusions}

We studied the problem of designing a dynamic payoff mechanism in epidemic population games. Unlike existing studies, such as \cite{MARTINS2023111016,certorio_epidemic_2022}, our work analyzes the dynamics underlying the games where the agents' decision making is subject to perturbation. 
We adopted (PBR EDM) to formalize this perturbed decision-making process and established stability of the feedback interconnection between (EPG) and (PBR EDM). Notably, our main results demonstrate the global convergence of the average transmission rate to its achievable minimum, subject to the budget constraint, thereby minimizing the infectious population in the long run. Additionally, we established an anytime bound on the infectious population using the Lyapunov stability method. We also discussed scenarios where the social planner needs to estimate the parameter of the agents' learning rule. 
As a direction for future research, we plan to analyze the transient behavior of the average cost trajectory. In the current study, we focused on designing (EPG) for the minimization of the average transmission rate while satisfying the asymptotic budget constraint. However, this consideration does not provide sufficient insight into the variation of the transient average cost. 

\bibliographystyle{elsarticle-num}        
\bibliography{MainRef}

\section*{Acknowledgement}
The authors would like to thank the anonymous reviewers for their constructive comments that led to substantial improvements in the paper.

\appendix

\subsection{Evaluation of (EPG) under Different Learning Rules for (PBR EDM)} \label{appendix_a}
\begin{figure}
    \centering
    \begin{subfigure}[b]{0.19\textwidth}
        \centering
        \ifnum \useTIKZ>0
            \resizebox{\textwidth}{!}{\input{figures/1.b.bottom.SIRS_EDM_I_ratio_0.15_nu2.0.tikz}}
        \else
            \includegraphics[height=4.5cm, width=\textwidth,trim={15mm 15mm 10mm 15mm},clip]{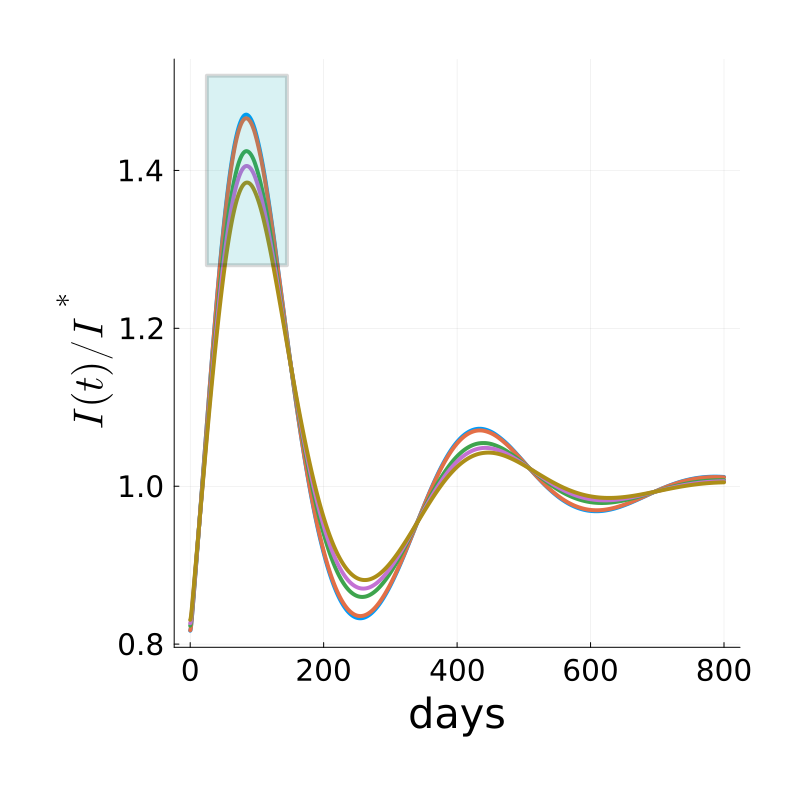}
        \fi
        \caption{}
        \label{fig:exampleManyDistributions:a}
    \end{subfigure}%
    \begin{subfigure}[b]{0.28\textwidth}
        \centering
        \ifnum \useTIKZ>0
            \resizebox{.8\textwidth}{!}{\input{figures/1.d.bottom.SIRS_EDM_I_ratio_peak_0.15_nu2.0.tikz}}
        \else
            \includegraphics[width=\textwidth,trim={15mm 15mm 10mm 15mm},clip]{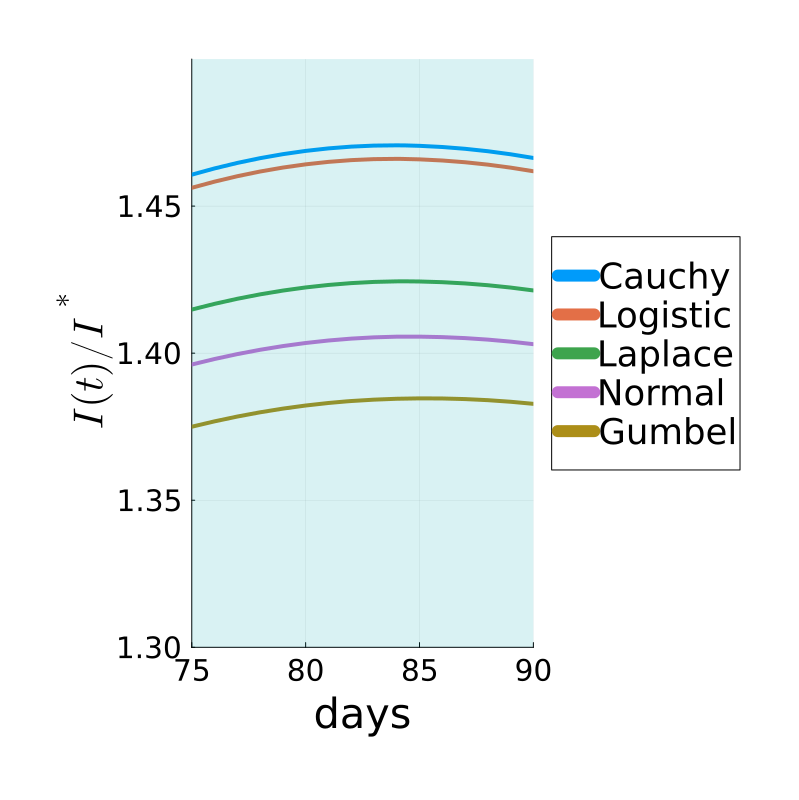}
        \fi
        \caption{}
        \label{fig:exampleManyDistributions:b}
    \end{subfigure}
    \caption{The ratio $I\st/I^\ast$ for (EPG) evaluated with different noise distributions for the choice function \eqref{eq:perturbed_revision_protocol}. The square area highlighted in (a) is presented at an enlarged scale in (b).}
    \label{fig:exampleManyDistributions}
\end{figure}

Fig.~\ref{fig:exampleManyDistributions} depicts the infectious fraction of the population when the probability distributions of the random variables $v_1, \cdots, v_n$ for \eqref{eq:perturbed_revision_protocol} are defined by the Cauchy, logistic, Laplace, normal, and generalized extreme value (Gumbel) distributions. From the figure, we can observe that the trajectory of the infectious fraction does not vary substantially, even when the choice function is defined by different noise distributions.

\subsection{Proof of Theorem~\ref{theorem:stability}} \label{sec:proof_stability_theorem}

To prove the theorem, we first introduce the following lemma.
\begin{lemma} \label{lemma:c_increasing_function}
    The function $\vec \beta' C(q \vec \beta + \bar r - \tilde c (\bar I))$ is increasing with respect to $q$.
\end{lemma}
\begin{proof}
    To proceed, we establish that $\frac{\partial}{\partial q} \vec \beta' C(q \vec \beta + \bar r - \tilde c (\bar I)) > 0, ~ \forall q \in \mathbb R$. Note that
\begin{align}
     & \frac{\partial}{\partial q} \vec \beta' C(q \vec \beta + \bar r - \tilde c (\bar I)) \nonumber                                                           \\
     & = \frac{\partial}{\partial q} \vec \beta' \argmax_{z \in \mathrm{int} (\mathbb X)} ( z' (q\vec{\beta} + \bar r - \tilde c (\bar I))  - Q (z) ) \nonumber \\
     & = \vec \beta' \nabla_p \argmax_{z \in \mathrm{int} (\mathbb X)} ( z' p - Q (z) ) \Big|_{p = q\vec{\beta} + \bar r - \tilde c (\bar I)} \vec \beta.
\end{align}
By the same arguments used in the proof of \cite[Theorem~2.1]{Hofbauer2002On-the-global-c}, we conclude that
\begin{align}
    \vec \beta' \nabla_p \argmax_{z \in \mathrm{int} (\mathbb X)} ( z' p - Q (z) ) \big|_{p = q\vec{\beta} + \bar r - \tilde c (\bar I)} \vec \beta > 0.
\end{align}
Therefore, $\vec \beta' C(q \vec\beta + \bar r - \tilde c (\bar I))$ is an increasing function of $q$ satisfying
\begin{subequations}
    \begin{align}
         & \lim_{q \to \infty} \vec \beta' C(q \vec\beta + \bar r - \tilde c (\bar I)) = \vec \beta_n   \\
         & \lim_{q \to -\infty} \vec \beta' C(q \vec\beta + \bar r - \tilde c (\bar I)) = \vec \beta_1.
    \end{align}
\end{subequations}
\end{proof}

We provide a three-part proof in which we establish $\lim_{t \to \infty} \|\dot x\st\|_2 = \lim_{t \to \infty} |I\st - \hat I\st| = \lim_{t \to \infty} |R\st - \hat R\st| = 0 $ in \textbf{Part~2} and $\lim_{t \to \infty} q\st = \bar q$ and $\lim_{t \to \infty} \mathcal B\st = \bar \beta$ in \textbf{Part~3}, where $\bar q$ is a unique solution to $\bar\beta = \vec\beta' C(\bar q \vec\beta + \bar r - \tilde c (\bar I))$. Consequently, in conjunction with $\lim_{t \to \infty} \|x\st - C(q\st \vec\beta + \bar r - \tilde c (\bar I))\|_2 = \lim_{t \to \infty} \|\dot x\st \|_2 = 0$, we can conclude that the two statements of the theorem are valid.

\textit{\textbf{Part~1.} $\mathcal I\st, \hat{\mathcal I}\st$ are strictly positive for all $t \geq 0$}: Adopting the same Lyapunov function $\mathscr S$ from \cite{MARTINS2023111016}, we can state
\begin{align} \label{eq:storage_function_for_EPG}
     & \mathscr S ( \mathcal I, \mathcal R, \mathcal B ) \nonumber                                                                                                                                                     \\
     & = ( \mathcal I - \hat{\mathcal I} ) + \hat{\mathcal I} \ln \frac{\hat{\mathcal I}}{\mathcal I} + \frac{1}{2 \gamma} ( \mathcal R - \hat{\mathcal R} )^2 + \frac{v^2}{2} ( \mathcal B - \bar \beta )^2 \nonumber \\
     & = ( \mathcal I - \eta ( \mathcal B - \sigma ) ) + \eta ( \mathcal B - \sigma ) \ln \frac{\eta ( \mathcal B - \sigma )}{\mathcal I} \nonumber                                                                    \\
     & \qquad + \frac{1}{2 \gamma} ( \mathcal R \!-\! (1 \!-\! \eta) ( \mathcal B \!-\! \sigma ) )^2 + \frac{v^2}{2} ( \mathcal B \!-\! \bar \beta )^2,
\end{align}
where we use $\hat{\mathcal I} = \mathcal B \hat I = \eta (\mathcal B - \sigma)$ and $\hat{\mathcal R} = \mathcal B \hat R = (1 - \eta) (\mathcal B - \sigma)$.
Note that using (EPG) and (PBR EDM), we can derive the following relation.
\begin{align} \label{eq:lyapunov_function_time_derivative}
     & \kappa^{-1} \frac{\mathrm d}{\mathrm dt} \mathcal S(x\st, p\st) + \frac{\mathrm d}{\mathrm dt} \mathscr S(\mathcal I\st, \mathcal R\st, \mathcal B\st) \nonumber \\
     & = \kappa^{-1} \nabla_x' \mathcal S(x\st, p\st ) \mathcal V(x\st, p\st) \nonumber \\
     & \qquad - (\mathcal I\st - \hat{\mathcal I}\st)^2 - \frac{\omega}{\gamma}(\mathcal R\st - \hat{\mathcal R}\st)^2 \leq 0,
\end{align}
which implies, since $\mathcal S$ and $\mathscr S$ are non-negative functions, $\mathscr S$ is a bounded function -- both below and above. In conjunction with the fact that $\hat {\mathcal I} = \eta (\mathcal B - \sigma) \geq \eta (\mathcal B_{\text{min}} - \sigma) > 0$, we conclude that $\mathcal I\st$ is strictly positive for all $t \geq 0$, i.e., there is $\delta > 0$ for which $\mathcal I\st \geq \delta, ~ \forall t \geq 0$ holds. Otherwise, $\eta (\mathcal B\st - \sigma) \ln \frac{\eta (\mathcal B\st - \sigma)}{\mathcal I\st}$ tends to infinity and so does $\mathscr S(\mathcal I\st, \mathcal R\st, \mathcal B\st)$ as $\mathcal I\st$ approaches zero, which contradicts the fact that $\mathscr S$ is a bounded function.

We remark that from (EPGc) and the definition of the mapping $G$, since both $\mathcal I\st$ and $\hat{\mathcal I}\st$ are strictly positive for all $t \geq 0$ and also upper bounded according to their respective definitions, we can infer that $\dot q \st$ is bounded.

\textit{\textbf{Part~2.} It holds that $\lim_{t \to \infty} \|\dot x\st\|_2 = \lim_{t \to \infty} |I\st - \hat I\st| = \lim_{t \to \infty} |R\st - \hat R\st| = 0 $}: 
For notational convenience, let us define
\begin{align} \label{eq:S_p_script}
    \breve{\mathscr S}(\mathcal I, \mathcal R, \mathcal B) = ( \mathcal I - \hat{\mathcal I} ) + \hat{\mathcal I} \ln \frac{\hat{\mathcal I}}{\mathcal I} + \frac{1}{2 \gamma} ( \mathcal R - \hat{\mathcal R} )^2.
\end{align}
Note that $0 \leq \breve{\mathscr S} (\mathcal I, \mathcal R, \mathcal B) \leq \mathscr S(\mathcal I, \mathcal R, \mathcal B)$. According to \eqref{eq:lyapunov_function_time_derivative} and \eqref{eq:S_p_script}, we can derive
\begin{align} \label{eq:lyapunov_function_inequality}
     & \kappa^{-1} \mathcal S(x\st, p\st) + \breve{\mathscr S} (\mathcal I\st, \mathcal R\st, \mathcal B\st) - \alpha \nonumber                                               \\
     & \leq \kappa^{-1} \mathcal S(x\st, p\st) + \mathscr S(\mathcal I\st, \mathcal R\st, \mathcal B\st) - \alpha \nonumber                                                   \\
     & =\int_0^t \big( \kappa^{-1} \nabla_x' \mathcal S(x\sta{\tau}, p\sta{\tau}) \mathcal V(x\sta{\tau}, p\sta{\tau}) \nonumber                                                               \\
     & \quad - (\mathcal I\sta{\tau} - \hat{\mathcal I}\sta{\tau})^2 - \frac{\omega}{\gamma}(\mathcal R\sta{\tau} - \hat{\mathcal R}\sta{\tau})^2 \big) \, \mathrm d \tau, ~ \forall t \geq 0,
\end{align}
where $\alpha = \kappa^{-1} \mathcal S(x\so, p\so) + \mathscr S(\mathcal I\so, \mathcal R\so, \mathcal B\so)$.  We claim that
\begin{align} \label{eq:storage_function_vanishes}
    \lim_{t \to \infty} \left( \kappa^{-1} \mathcal S(x\st, p\st) + \breve{\mathscr S} (\mathcal I\st, \mathcal R\st, \mathcal B\st) \right) = 0,
\end{align}
which implies
\begin{align*}
    \lim_{t \to \infty} \mathcal S(x\st, p\st) & = 0                                                                                        \\
    \lim_{t \to \infty} |I\st - \hat I\st|     & \leq \lim_{t \to \infty} \frac{1}{\vec \beta_1} |\mathcal I\st - \hat{\mathcal I}\st| = 0  \\
    \lim_{t \to \infty} |R\st - \hat R\st|     & \leq \lim_{t \to \infty} \frac{1}{\vec \beta_1} |\mathcal R\st - \hat{\mathcal R}\st| = 0.
\end{align*}
Furthermore, according to the analysis used in \cite[Theorem~2.1]{Hofbauer2002On-the-global-c},
\begin{align} \label{eq:legendre_transformation_equivalence}
    \tilde z' \nabla Q(y) = \tilde z'r \!\iff\! y = \argmax_{z \in \mathrm{int} ( \mathbb X )} ( z'r - Q(z) )
\end{align}
holds for all $r \in \mathbb R^n$, $y \in \mathrm{int}(\mathbb X)$, and $\tilde z \in T\mathbb X$. Let $y\st = \argmax_{z \in \mathrm{int} ( \mathbb X )} ( z' p\st - Q(z) )$. Then, we can derive
\begin{multline} \label{eq:storage_function_convexity}
    \mathcal S(x\st, p\st) = (y\st - x\st)' \nabla Q(y\st) \\
    - (Q (y\st) - Q (x\st)).
\end{multline}
Therefore, by the strict convexity \eqref{eq:perturbation_conditions_a} of $Q$ and \eqref{eq:storage_function_convexity}, $\lim_{t \to \infty} \mathcal S(x\st, p\st) = 0$ implies that $\lim_{t \to \infty} \|\dot x\st\|_2 = 0$.

In what follows, we justify the claim. Let us define the following set $\mathbb O_\epsilon$:
\begin{align*}
    \mathbb O_\epsilon = \Big\{ t > 0 \,\Big|\, \kappa^{-1} \mathcal S(x\st, p\st) + \breve{\mathscr S} (\mathcal I\st, \mathcal R\st, \mathcal B\st) > \frac{\epsilon}{2} \Big\}.
\end{align*}
Since $\mathcal S$ and $\breve{\mathscr S}$ are differentiable, $\mathbb O_\epsilon$ is an open set and can be represented as an infinite union of disjoint open intervals, i.e., $\mathbb O_\epsilon = \cup_{i=1}^\infty \mathbb I_i$ where $\mathbb I_i = (a_i, b_i)$ is an open interval.  Note that
\begin{align}
    \kappa^{-1} \mathcal S(x(a_i), p(a_i)) + \breve{\mathscr S} (\mathcal I(a_i), \mathcal R(a_i), \mathcal B(a_i)) = \frac{\epsilon}{2}.
\end{align}
The $\delta$-storage function of (PBR EDM) satisfies
\begin{align} \label{eq:storage_function_upper_bound}
    \mathcal S(x, p) & = y' p  - Q(y) - (x'p - Q(x)) \nonumber                  \\
                     & = p' (y-x) - (Q(y) - Q(x)) \nonumber                     \\
                     & \overset{(i)}{\leq} (p - \nabla Q(x))' (y - x) \nonumber \\
                     & = - \nabla_x' \mathcal S(x, p) \mathcal V(x, p),
\end{align}
where $y = \argmax_{z \in \mathrm{int} ( \mathbb X )} ( z' p - Q(z) )$. For $(i)$, we use the convexity \eqref{eq:perturbation_conditions_a} of $Q$ to establish $Q(y) - Q(x) \geq \nabla' Q(x) (y-x)$.
Also, by \textbf{Part~1}, since $\mathcal I, \hat{\mathcal I}$ are strictly positive, i.e., there is $\delta>0$ for which $\mathcal I, \hat{\mathcal I} \geq \delta$ holds, we can find a positive constant $k$ satisfying the inequality given by
\begin{align} \label{eq:inequality_I_minus_I_hat}
    k | \mathcal I - \hat{\mathcal I} | \geq (\mathcal I - \hat{\mathcal I}) + \hat{\mathcal I} \ln \frac{\hat{\mathcal I}}{\mathcal I}.
\end{align}
For instance, \eqref{eq:inequality_I_minus_I_hat} holds with $k$ defined as
\begin{align*}
    k = 1 + \eta(\vec \beta_n - \sigma) \max_{\mathcal I, \hat{\mathcal I} \geq \delta} \frac{\ln \hat{\mathcal I} - \ln \mathcal I}{\hat{\mathcal I} - \mathcal I}.
\end{align*}

Consequently, by \eqref{eq:storage_function_upper_bound} and \eqref{eq:inequality_I_minus_I_hat}, for every $t \in \mathbb O_\epsilon$, there is a constant $\delta_{\epsilon} > 0$ for which
\begin{multline} \label{eq:derivative_strictly_negative}
    \kappa^{-1} \nabla_x' \mathcal S(x\st, p\st) \mathcal V(x\st, p\st) \\ -
    (\mathcal I\st - \hat{\mathcal I}\st)^2 - \frac{\omega}{\gamma}(\mathcal R\st - \hat{\mathcal R}\st)^2 < -\delta_{\epsilon}.
\end{multline}

Since $\kappa^{-1} \mathcal S(x\st, p\st) + \breve{\mathscr S} (\mathcal I\st, \mathcal R\st, \mathcal B\st)$ is a non-negative function of $t$, from \eqref{eq:lyapunov_function_inequality} and \eqref{eq:derivative_strictly_negative}, we can derive
\begin{align}
    -\alpha & \leq \int_0^\infty \big( \kappa^{-1} \nabla_x' \mathcal S(x\sta{\tau}, p\sta{\tau}) \mathcal V(x\sta{\tau}, p\sta{\tau}) \nonumber                                                        \\
            & \qquad - (\mathcal I\sta{\tau} - \hat{\mathcal I}\sta{\tau})^2 - \frac{\omega}{\gamma}(\mathcal R\sta{\tau} - \hat{\mathcal R}\sta{\tau})^2 \big) \, \mathrm d \tau \nonumber \\
            & \leq \int_{\mathbb O_\epsilon} \big( \kappa^{-1} \nabla_x' \mathcal S(x\sta{\tau}, p\sta{\tau}) \mathcal V(x\sta{\tau}, p\sta{\tau}) \nonumber                                            \\
            & \qquad - (\mathcal I\sta{\tau} - \hat{\mathcal I}\sta{\tau})^2 - \frac{\omega}{\gamma}(\mathcal R\sta{\tau} - \hat{\mathcal R}\sta{\tau})^2 \big) \, \mathrm d \tau \nonumber \\
            & \leq - \delta_{\epsilon} \mathcal L(\mathbb O_\epsilon),
\end{align}
where $\mathcal L(\mathbb O_\epsilon)$ is the Lebesgue measure of $\mathbb O_\epsilon$.
Therefore, the set $\mathbb O_\epsilon$ has finite Lebesgue measure which implies that $\lim_{i \to \infty} |b_i - a_i| = 0$.

To complete the proof of the claim, we argue that for every $\epsilon > 0$, there is $T_{\epsilon} > 0$ for which it holds that
\begin{multline}
    \kappa^{-1} \mathcal S(x\st, p\st) + \breve{\mathscr S} (\mathcal I\st, \mathcal R\st, \mathcal B\st) < \epsilon, ~ \forall t \geq T_\epsilon.
\end{multline}
If the argument does not hold, then for some $\epsilon > 0$, we can find an infinite subset $\mathbb J$ of $\mathbb N$ for which
\begin{align} \label{eq:lyapunov_function_positive_assumption}
    \max_{t \in \bar{\mathbb I}_j} \left( \kappa^{-1} \mathcal S(x\st, p\st) + \breve{\mathscr S} (\mathcal I\st, \mathcal R\st, \mathcal B\st) \right) \geq \epsilon
\end{align}
holds for all $j$ in $\mathbb J$, where $\bar{\mathbb I}_j = [a_j, b_j]$ is the closure of the open subset $\mathbb I_j = (a_j, b_j)$ of $\mathbb O_\epsilon$. Let $\bar t_j$ be a time instant in $\bar{\mathbb I}_j$ attaining $\max_{t \in \bar{\mathbb I}_j} ( \kappa^{-1} \mathcal S(x\st, p\st) + \breve{\mathscr S} (\mathcal I\st, \mathcal R\st, \mathcal B\st) )$. We can derive the following relation.
\begin{align}
     & \kappa^{-1} \mathcal S(x\sta{\bar t_j}, p\sta{\bar t_j}) + \breve{\mathscr S} (\mathcal I\sta{\bar t_j}, \mathcal R\sta{\bar t_j}, \mathcal B\sta{\bar t_j}) \nonumber                                                                     \\
     & \qquad - \kappa^{-1} \mathcal S(x\sta{a_j}, p\sta{a_j}) - \breve{\mathscr S} (\mathcal I\sta{a_j}, \mathcal R\sta{a_j}, \mathcal B\sta{a_j}) \nonumber                                                                                     \\
     & = \int_{a_j}^{\bar t_j} \frac{\mathrm d}{\mathrm d\tau} \left( \kappa^{-1} \mathcal S(x\sta{\tau}, p\sta{\tau}) + \breve{\mathscr S} (\mathcal I\sta{\tau}, \mathcal R\sta{\tau}, \mathcal B\sta{\tau}) \right) \, \mathrm d\tau \nonumber \\
     & \overset{(i)}{\leq} -\int_{a_j}^{\bar t_j} v^2 \dot{\mathcal B}\sta{\tau} (\mathcal B\sta{\tau} - \bar \beta) \, \mathrm d \tau \nonumber                                                                          \\
     & \leq \int_{a_j}^{\bar t_j} v^2 \|\vec \beta \|_2 \|\dot x \sta{\tau} \|_2 (\|\vec \beta \|_2 \|x \sta{\tau} \|_2 + \bar \beta) \, \mathrm d \tau.
\end{align}
To obtain $(i)$, we use \eqref{eq:lyapunov_function_time_derivative} and the definition of $\breve{\mathscr S}$.
Since both $\dot x$ and $x$ are bounded, we can find a constant $M$ for which the following inequality holds.
\begin{align}
     & \kappa^{-1} \mathcal S(x\sta{\bar t_j}, p\sta{\bar t_j}) + \breve{\mathscr S} (\mathcal I\sta{\bar t_j}, \mathcal R\sta{\bar t_j}, \mathcal B\sta{\bar t_j}) \nonumber \\
     & \qquad - \kappa^{-1} \mathcal S(x\sta{a_j}, p\sta{a_j}) - \breve{\mathscr S} (\mathcal I\sta{a_j}, \mathcal R\sta{a_j}, \mathcal B\sta{a_j}) \nonumber                 \\
     & \leq M (\bar t_j - a_j).
\end{align}
On the other hand, since $\lim_{j \to \infty} |\bar t_j - a_j| \leq \lim_{j \to \infty} |b_j - a_j| = 0$, for sufficiently large index $j$ in $\mathbb J$, we have that
\begin{align*}
     & \kappa^{-1} \mathcal S(x\sta{\bar t_j}, p\sta{\bar t_j}) + \breve{\mathscr S} (\mathcal I\sta{\bar t_j}, \mathcal R\sta{\bar t_j}, \mathcal B\sta{\bar t_j}) \nonumber \\
     & < \kappa^{-1} \mathcal S(x\sta{a_j}, p\sta{a_j}) + \breve{\mathscr S} (\mathcal I\sta{a_j}, \mathcal R\sta{a_j}, \mathcal B\sta{a_j}) + \frac{\epsilon}{2} = \epsilon,
\end{align*}
which contradicts \eqref{eq:lyapunov_function_positive_assumption}. This completes \textbf{Part~2}.

\textit{\textbf{Part~3.} It holds that $\lim_{t \to \infty} q\st = \bar q$ and $\lim_{t \to \infty} \mathcal B\st = \bar \beta$}: 
\begin{figure*}[t]
    \begin{multline} \label{eq:q_dynamics_pbr}
        \dot q\st = \kappa v^2 \big(
        \bar \beta - \vec \beta' C(q\st \vec\beta + \bar r - \tilde c (\bar I)) \big) \\
        + \underbrace{\kappa v^2 \big( \vec \beta' C(q\st \vec\beta + \bar r - \tilde c (\bar I)) - \mathcal B\st \big)
        + \kappa ( \hat I\st - I\st ) + \kappa \eta ( \ln I\st - \ln \hat I\st ) + \kappa \frac{\mathcal B\st}{\gamma} ( R\st - \hat R\st ) ( 1 - \eta - R\st )}_{=\epsilon\st}
    \end{multline}
\change{
\begin{align} \label{eq:linearization}
  \begin{pmatrix} \dot{\tilde I}\st \\ \dot{\tilde R}\st \\ \dot{\tilde q}\st \\ \dot{\tilde x}\st \end{pmatrix} =
  \begin{pmatrix}
    - \eta (\bar \beta - \sigma) & - \eta (\bar \beta - \sigma) & 0 & 
    \eta \left( 1 - \frac{\sigma}{\bar \beta} \right) \frac{\sigma}{\bar \beta} \vec \beta' \\
    \gamma & -\omega & 0 & 0 \\
    \frac{\kappa \sigma}{\bar \beta - \sigma} & \frac{\kappa \sigma}{\gamma} (1 - \eta) & 0 &  -\frac{\kappa \sigma^2}{\bar \beta^2} \left(  \frac{\eta}{\bar \beta - \sigma} 
      + \frac{\left( 1 - \eta \right)^2}{\gamma } \right) \vec \beta' - \kappa \upsilon^2 \vec \beta' \\
    0 & 0 & d & - I_n
  \end{pmatrix} \begin{pmatrix} \tilde I \st \\ \tilde R\st \\ \tilde q\st \\ \tilde x\st \end{pmatrix}
\end{align}
\begin{align} \label{eq:linearization_reduced}
  \begin{pmatrix} \dot{\tilde I}\st \\ \dot{\tilde R}\st \\ \dot{\tilde q}\st \\ \dot{\tilde{\mathcal B}}\st \end{pmatrix} =
  \underbrace{\begin{pmatrix}
    - \eta (\bar \beta - \sigma) & - \eta (\bar \beta - \sigma) & 0 & 
    \eta \left( 1 - \frac{\sigma}{\bar \beta} \right) \frac{\sigma}{\bar \beta} \\
    \gamma & -\omega & 0 & 0 \\
    \frac{\kappa \sigma}{\bar \beta - \sigma} & \frac{\kappa \sigma}{\gamma} (1 - \eta) & 0 &  -\frac{\kappa \sigma^2}{\bar \beta^2} \left(  \frac{\eta}{\bar \beta - \sigma} 
      + \frac{\left( 1 - \eta \right)^2}{\gamma } \right) - \kappa \upsilon^2  \\
    0 & 0 & \vec\beta' d & - 1
  \end{pmatrix}}_{=A} \begin{pmatrix} \tilde I \st \\ \tilde R\st \\ \tilde q\st \\ \tilde{\mathcal B}\st \end{pmatrix},
\end{align}
where $\tilde I\st = I\st - \bar I$, $\tilde R\st = R\st - \bar R$, $\tilde q\st = q\st - \bar q$, $\tilde x\st = x\st - \bar x$, and $d = (d_1 ~ \cdots ~ d_n)'$ with $d_i = \bar x_i ( \vec \beta_i - \bar \beta )/\mu$.
\begin{multline} \label{eq:characteristic_polynomial}
    \mathrm{det} (A - \lambda I) 
    = \lambda^4 + \left( 1 + \omega + \eta (\bar \beta - \sigma) \right) \lambda^3 + \left( \omega + (\eta + \omega) (\bar \beta - \sigma) + \vec \beta' d \left( \frac{\kappa \sigma^2}{\bar \beta^2} \left(  \frac{\eta}{\bar \beta - \sigma} + \frac{\left( 1 - \eta \right)^2}{\gamma } \right) + \kappa \upsilon^2 \right) \right) \lambda^2 \\
    + \left( \omega (\bar \beta - \sigma) + \kappa (\omega + \eta (\bar \beta - \sigma)) \vec \beta' d \upsilon^2 + \omega \vec \beta' d \frac{\kappa \sigma^2}{\bar \beta^2} \frac{\eta}{\bar \beta - \sigma} + \eta (\bar \beta - \sigma) \vec \beta' d \frac{\kappa \sigma^2}{\bar \beta^2} \frac{\left( 1 - \eta \right)^2}{\gamma } \right) \lambda + \kappa \vec \beta' d \omega (\bar \beta - \sigma) \upsilon^2 = 0.
\end{multline}
\hrulefill
}
\end{figure*}
Since $p\st = r\st - \tilde c (I\st) = q\st \vec \beta + \bar r - \tilde c (\bar I)$, we can re-write (EPGc) as in \eqref{eq:q_dynamics_pbr}. By \textbf{Part~2}, we observe that
\begin{align*}
     & \lim_{t \to \infty} \big( \vec \beta' C(q\st \vec\beta + \bar r - \tilde c (\bar I)) - \mathcal B\st \big) = \lim_{t \to \infty} \vec \beta' \dot x\st = 0 \\
     & \lim_{t \to \infty} |I\st - \hat I\st| = 0                                                                                                        \\
     & \lim_{t \to \infty} |R\st - \hat R\st| = 0,
\end{align*}
which, in conjunction with the fact that $\textstyle\max_{\mathcal I, \hat{\mathcal I} \geq \delta} \frac{\ln \hat{\mathcal I} - \ln \mathcal I}{\hat{\mathcal I} - \mathcal I} > 0$ is finite, imply that $\lim_{t \to \infty} \epsilon\st = 0$, where $\epsilon\st$ is defined in \eqref{eq:q_dynamics_pbr}. Hence, if $q\st$ goes to $\bar q$, satisfying $\bar \beta = \vec \beta' C(\bar q \vec\beta + \bar r - \tilde c (\bar I))$, as $t$ tends to infinity, then we can infer that $\lim_{t \to \infty} \mathcal B\st = \vec \beta' C(\bar q \vec\beta + \bar r - \tilde c (\bar I)) = \bar \beta$.

Using Lemma~\ref{lemma:c_increasing_function}, we can infer that there is $\bar q$ for which $\bar \beta = \vec \beta' C(\bar q \vec\beta + \bar r - \tilde c (\bar I))$ holds and together with $\lim_{t \to \infty} \epsilon\st = 0$, $q\st$ converges to $\bar q$ as $t$ tends to infinity. In particular, if $\bar \beta = \vec \beta' C(\bar r - \tilde c (\bar I))$ then $\bar q = 0$. This completes the proof. \QED

\subsection{Proof of Proposition~\ref{prop:local_exponential_stability}} \label{sec:proof_local_exponential_stability}

\change{
We first prove the Lyapunov stability of the closed-loop model. To this end, we show that 
\begin{align}
    \kappa^{-1} \mathcal S(x, p) + \mathscr S (\mathcal I, \mathcal R, \mathcal B) = 0
\end{align}
holds if and only if
\begin{subequations} \label{eq:lyapunov_stability_condition}
\begin{align}
    I &= \bar I \label{eq:lyapunov_stability_condition_a} \\
    R &= \bar R \label{eq:lyapunov_stability_condition_b} \\
    q &= \bar q \label{eq:lyapunov_stability_condition_c} \\
    x &= \bar x, \label{eq:lyapunov_stability_condition_d}
\end{align}
\end{subequations}
hold, where $\bar x = C (\bar q \vec\beta + \bar r - \tilde c(\bar I))$. Since the ``if'' part directly follows from the definitions of $\mathcal S$ and $\mathscr S$, we discuss the proof of the ``only if'' part.

Eqs. \eqref{eq:lyapunov_stability_condition_a} and \eqref{eq:lyapunov_stability_condition_b} follow from the definition of $\mathscr S$. Also, from the definition of $\mathscr S$, we can infer that $\mathcal B = \bar \beta$. Recall that, by the strict convexity \eqref{eq:perturbation_conditions_a} of $Q$ and \eqref{eq:storage_function_convexity}, $\mathcal S(x, p) = 0$ implies $C(q \vec\beta + \bar r - \tilde c(\bar I)) = x$. Consequently, by the fact that $\vec\beta' C(q \vec\beta + \bar r - \tilde c(\bar I))$ is an increasing function of $q$ as shown in Lemma~\ref{lemma:c_increasing_function},
we can infer that \eqref{eq:lyapunov_stability_condition_c} and \eqref{eq:lyapunov_stability_condition_d} hold. 
As verified in the proof of Theorem~\ref{theorem:stability}, $\frac{\mathrm d}{\mathrm dt} (\kappa^{-1} \mathcal S(x\st, p\st) + \mathscr S (\mathcal I\st, \mathcal R\st, \mathcal B\st)) \leq 0$ holds along the solution trajectory of the closed-loop model. Therefore, we conclude that the closed-loop model is Lyapunov stable. Additionally, in conjunction with the results of Theorem~\ref{theorem:stability}, we further conclude that the closed-loop model is asymptotically stable.

By linearizing the closed-loop model around $(\bar I, \bar R, \bar q, \bar x)$, we can derive \eqref{eq:linearization}. Note that all eigenvalues of \eqref{eq:linearization} have negative real parts if the reduced model \eqref{eq:linearization_reduced} also exhibits this property. To conclude local exponential stability, given the model's asymptotic stability, it is sufficient to show that \eqref{eq:linearization_reduced} does not have any eigenvalues on the imaginary axis.

Consider the characteristic polynomial \eqref{eq:characteristic_polynomial} for \eqref{eq:linearization_reduced}. When $\upsilon = 0$, \eqref{eq:characteristic_polynomial} has one solution at the origin, and we claim that the remaining three non-zero solutions are not on the imaginary axis. Conversely, when $\upsilon > 0$, \eqref{eq:characteristic_polynomial} has no solution at the origin. Since the solutions of \eqref{eq:characteristic_polynomial} continuously depend on $\upsilon$, we conclude that all the eigenvalues of \eqref{eq:linearization_reduced} are off the imaginary axis for small $\upsilon > 0$.
To prove the claim, consider $\lambda = j b$, where $b$ is a nonzero real number. By substituting $\lambda = j b$ in  \eqref{eq:characteristic_polynomial} with $\upsilon=0$, we obtain the following two equations:
\begin{subequations}
\begin{align} \label{eq:imaginary_root_1}
  b^2 \!=\! \Bigg( \omega \!+\! (\eta \!+\! \omega) (\bar \beta \!-\! \sigma) \!+\! \vec \beta' d \frac{\kappa \sigma^2}{\bar \beta^2} \left(  \frac{\eta}{\bar \beta - \sigma} \!+\! \frac{\left( 1 - \eta \right)^2}{\gamma } \right)  \Bigg) 
\end{align}
\vspace{-1.2em}
\begin{multline} \label{eq:imaginary_root_2}
  \left( 1 + \omega + \eta (\bar \beta - \sigma) \right) b^2
  = \omega (\bar \beta - \sigma)
  + \omega \vec \beta' d \frac{\kappa \sigma^2}{\bar \beta^2} \frac{\eta}{\bar \beta - \sigma} \\
  + \eta (\bar \beta - \sigma) \vec \beta' d \frac{\kappa \sigma^2}{\bar \beta^2} \frac{\left( 1 - \eta \right)^2}{\gamma }.
\end{multline}
\end{subequations}
Note that $\vec\beta' d \geq 0$.
However, from \eqref{eq:imaginary_root_1}, we can infer that
\begin{multline*}
  \left( 1 + \omega + \eta (\bar \beta - \sigma) \right) b^2
  > \omega (\bar \beta - \sigma)
  + \omega \vec \beta' d \frac{\kappa \sigma^2}{\bar \beta^2} \frac{\eta}{\bar \beta - \sigma} \\
  + \eta (\bar \beta - \sigma) \vec \beta' d \frac{\kappa \sigma^2}{\bar \beta^2} \frac{\left( 1 - \eta \right)^2}{\gamma },
\end{multline*}
which contradicts \eqref{eq:imaginary_root_2}. 
Consequently, we conclude no eigenvalues lie on the imaginary axis. \QED
}

\subsection{Proof of Lemma~\ref{proposition:estimated_perturbation_parameter}} \label{sec:proof_proposition_estimated_perturbation_parameter}
Noting that 
\begin{align*}
    &(r - \tilde c (I))' C^{\mu}(r - \tilde c (I)) \\
    &= (r \!-\! \tilde c (I))' \argmax_{z \in \mathrm{int} (\mathbb X)} ( \lambda z' (r \!-\! \tilde c (I))\!-\! \bar Q (z) ),
\end{align*}
where $\lambda = \mu^{-1}$, the proof of this lemma is similar to that of Lemma~\ref{lemma:c_increasing_function}. We omit details for brevity. \QED

\change{
\subsection{Proof of Proposition~\ref{prop:parameter_estimation}}
With $r = \tilde r + \tilde c(I)$, let $\mathbf R$ be a discrete random variable whose probability is defined as $\mathbb P (\mathbf R = \tilde r_i) = C_i^\mu (\tilde r)$. Note that $\mathbb E [\mathbf R] = \tilde r' C^\mu (\tilde r)$ holds. After surveying $K$ randomly selected agents, using the answers collected, we can compute $K$ realizations $\{R^{(k)}\}_{k=1}^K$ of $\mathbf R$ with $R^{(k)} = \tilde r' \xi_k$. Then, by applying Chebyshev's inequality, we can establish
\begin{align} \label{eq:upper_bound_of_ER}
    \mathbb P \left( \left| \mathbb E[\mathbf R] - \frac{1}{K} \textstyle \sum_{k=1}^K R^{(k)} \right| \leq \epsilon \right) 
    & \geq 1 - 1 / \epsilon^2 K,
\end{align}
where, to establish the inequality, we use \eqref{eq:condition_for_popoviciu_inequlity} and Popoviciu's inequality. The inequality \eqref{eq:upper_bound_of_ER} explains how to find both an upper and a lower bounds of $\tilde r' C^\mu (\tilde r)$ using the sample mean, with an arbitrarily high level of confidence. Consequently, using Lemma~\ref{proposition:estimated_perturbation_parameter}, we can compute bounds $\mu_L, \mu_R$ on $\mu$ as in \eqref{eq:bounds_on_mu}, leading to \eqref{eq:probabilistic_bounds_on_mu}.
\QED
}

\subsection{Proof of Proposition~\ref{proposition:cost_upper_bound}} \label{sec:proof_proposition_cost_upper_bound}

Since the budget constraint (P2) stated in {\bf Main Problem} is relevant only when the limit of reward $r\st$ has all non-negative entries, as explained in Remark~\ref{remark:choice_of_H}, we proceed with using \eqref{eq:H_nonnegative_incentive} for the definition of $H$ in (EPGd). According to Theorem~\ref{theorem:stability}, with $\bar r = \tilde c (\bar I)$, $q\st$ converges to $\bar q$ for which $\bar \beta = \vec \beta' C^\mu(\bar q \vec\beta)$ holds and, hence, $r\st$ converges to $\bar q \vec \beta + \tilde c (\bar I)$.

By defining $\lambda = \mu^{-1} \bar q$, we can express
\begin{align} \label{eq:choice_function_with_c}
    \bar \beta = \vec\beta' C^{\mu}(\bar q \vec\beta) = \vec\beta' \argmax_{z \in \mathrm{int} (\mathbb X)} ( z' (\lambda \vec\beta) - \bar Q (z) ).
\end{align}
By the same argument used in the proof of Lemma~\ref{proposition:estimated_perturbation_parameter}, $\vec\beta' \argmax_{z \in \mathrm{int} (\mathbb X)} ( z' (\lambda \vec\beta) - \bar Q (z) )$ is an increasing function of $\lambda$. Also, if $\bar \beta$ satisfies $\bar \beta < \vec\beta' C^{1} \so$ then 
it holds that $\lambda < 0$ and $\bar q < 0$.

When (EPG) and (PBR EDM) reach their equilibrium state, the planner would spend
\begin{align}
     & (\bar q \vec\beta + \tilde c (\bar I) - \bar q \vec\beta_n \mathbf 1 )' \argmax_{z \in \mathrm{int} (\mathbb X)} ( z' (\bar q  \vec\beta) - \mu \bar Q (z) ) \nonumber \\
     & =\mu \lambda (\bar \beta - \vec\beta_n) + \tilde c' (\bar I) \argmax_{z \in \mathrm{int} (\mathbb X)} ( z' (\lambda \vec\beta) - \bar Q (z) ),
\end{align}
which is upper bounded by
\begin{align}
    \mu_U \lambda (\bar \beta - \vec\beta_n) + \tilde c' (\bar I) \argmax_{z \in \mathrm{int} (\mathbb X)} ( z' (\lambda \vec\beta) - \bar Q (z) ).
\end{align}
This completes the proof.
\QED
\vspace{-4em}

\end{document}